\documentclass[review]{elsarticle}

\usepackage{lineno,hyperref}
\usepackage{latexsym}
\usepackage{amsfonts}
\usepackage{exscale}
\usepackage{amsbsy}
\usepackage{amsmath}
\usepackage{mathrsfs}
\usepackage{amssymb}
\usepackage{graphicx}
\usepackage{xcolor}
\usepackage{color}
 \usepackage{multirow}
\usepackage{geometry}
\usepackage{pdflscape}
\journal{Journal of Network and Computer Applications}

\begin{document}

\begin{frontmatter}
\title{Machine Learning for Intelligent Optical Networks: A Comprehensive Survey}

%
%
%
%

\author[mymainaddress]{Rentao Gu\corref{mycorrespondingauthor}}
\cortext[mycorrespondingauthor]{Corresponding author (E-mail: rentaogu@bupt.edu.cn)}
\author[mymainaddress]{Zeyuan Yang}
\author[mymainaddress]{Yuefeng Ji}
\address[mymainaddress]{Beijing Laboratory of Advanced Information Network,\\ Beijing University of Posts and Telecommunications (BUPT), Beijing, 100876, China.\\ \rm{\{rentaogu, buptyzy, jyf\}@bupt.edu.cn}}
\begin{abstract}

With the rapid development of Internet and communication systems, both in services and technologies, communication networks have been suffering increasing complexity. It is imperative to improve intelligence in communication network, and several aspects have been incorporating with Artificial Intelligence (AI) and Machine Learning (ML). Optical network, which plays an important role both in core and access network in communication networks, also faces great challenges of system complexity and the requirement of manual operations. To overcome the current limitations and address the issues of future optical networks, it is essential to deploy more intelligence capability to enable autonomous and flexible network operations. ML techniques are proved to have superiority on solving complex problems; and thus recently, ML techniques have been used for many optical network applications.

In this paper, a detailed survey of existing applications of ML for intelligent optical networks is presented. The applications of ML are classified in terms of their use cases, which are categorized into optical network control and resource management, and optical networks monitoring and survivability. The use cases are analyzed and compared according to the used ML techniques. Besides, a tutorial for ML applications is provided from the aspects of the introduction of common ML algorithms, paradigms of ML, and motivations of applying ML. Lastly, challenges and possible solutions of ML application in optical networks are also discussed, which intends to inspire future innovations in leveraging ML to build intelligent optical networks.

\end{abstract}


\end{frontmatter}

\section{Introduction}
Recently, with the rapid development of communication technologies (such as 5G, Internet of Things, and cloud computing), and emerging network services (such as VR/AR and 4K video), the data traffic in communication networks is growing exponentially. It is reported that, the global IP traffic will increase threefold from 2017 to 2022, and the number of devices connected to IP networks will be more than three times the global population by 2022 \cite{CISCO-VNT}. In response to the exponential growth of network traffic and increasing network operations complexity, it is imperative to enhance automation and intelligence of communication networks. 

Introducing intelligence into communication networks to improve network performance and facilitate network operations has become a hot topic both in industry and academia. From the industry aspect, Self-Driven Networks and Intent-Driven Networks are proposed, and have been put into extensive investigation. From the aspect of standardization process, ETSI establishes an Industry Specification Group named Experiential Networked Intelligence (ENI) to optimize and adjust the operator experience with the aid of Artificial Intelligence (AI) and Machine Learning (ML) \cite{ENI} \cite{ENI-WHITE}. ITU-T has established a focus group on ML for Future Networks including 5G (FG-ML5G) in 2017 \cite{ML5G}, and delivers a technical specification on unified architecture 5G and future networks with ML \cite{ML5G-ARCHITECTURE}. In the academic domain, AI, especially ML, has been used to provide intelligence in various communication systems and tasks, such as computer networks \cite{MLN}, cognitive radios \cite{MLCR}, Software Defined Networking (SDN) \cite{MLSDN}, wireless networks \cite{DLWN}, Wireless Sensor Network (WSN) \cite{MLWSN}, Internet of Things (IoT) \cite{DLIOT}, and traffic classification \cite{MLTC}. 

Especially, there is also a trend to introduce intelligence into optical networks for its fundamental role in the communication networks and its inherent complexity. Optical networks have been widely used as the major carrier network for traffic for several advantages, include wide bandwidth, low latency, and high anti-interference capability. It connects the upper layer services and the underlying physical resources: on one hand, it needs to provision the bandwidth for different service needs; and on the other hand, optical network involves resources allocation problem in multiple dimensions, such as wavelength, spectrum slot, and time slot. This situation makes optical network operation and maintenance more complicated than other communication networks.  If consider the convergence of optical network and other networks (such as 5G mobile network, and IP networks), it will become more serious. In general, there are mainly three challenges faced by development and operation of optical networks: 
\begin{itemize}
\item \textbf{Network complexity:} The number and complexity of optical network devices increase with the scale of optical networks extend. Additionally, the optical network acts as the bearer network in communication systems, which may carry multiple heterogeneous networks, such as 5G mobile networks, IoT, vehicle networking, and cloud computing. How to adapt the traffic from these different networks is becoming the first challenge for optical network operation and management.
\item \textbf{Service complexity:} Different Quality of Service (QoS) agreements require optical networks to provide differentiated services. And the novel techniques and applications, such as network slicing, ask the service provision to be implemented in real time. Then, the optical networks should furnish a flexible service providing mechanism to provide different levels of QoS to different services in the physical domain, which is difficult for a large network. 
\item \textbf{Resource management complexity:} The optical network is the bridge between upper layer traffic and underlying physical layer resources. So, it is responsible for physical layer resources allocation for traffic provision. However, there are multiple dimensional physical resources to be allocated, such as fiber, wavelength, spectrum, modulation format, and time slots. The joint assignment of multiple resources is time-consuming with high computational complexity.
\end{itemize}

With the increasing complexity, traditional manual operation in optical networks requires too much time, lacks the ability to handle such a complex and large-scale optical network, and may result in local optimization instead of global optimization \cite{JI_3C}. Thus, conventional control and management approaches cannot satisfy the target in low latency, scalability, and accuracy for future optical networks. To meet the demands of future optical networks, more intelligence should be introduced into the optical networks monitoring, control, and management to minimize manual intervention as well as increase network flexibility and automation level.

ML algorithms can deal with complex problems by iteratively learning from input data and environment feedback. Deploying ML to optical networks is a promising way to introduce intelligence into optical networks. Instead of manual-based, rule-based and static programming-based networks operations, intelligent optical network aided by ML techniques can learn inner relationship from data and environment to perform more automated and flexible network operations. 

Progresses have been made in applying AI and ML in optical networks, and several previous works have reviewed the researches in this field. Khan \emph{et al.} put emphasis on discussing supervised learning and unsupervised learning algorithms as well as their applications in transmission and Digital Signal Processing (DSP) problems in optical communications \cite{AOCP}. In \cite{AION}, the use cases of AI techniques in optical communications are surveyed. The paper discusses a wide range of intelligence algorithms, such as Genetic Algorithms (GA) and Ant Colony Optimization (ACO), which applied in optical communications. However, ML applications are only briefly introduced, which will be focused in our paper. In \cite{JOCN-MLNA}, the workflow and evaluation metrics of ML models, data and network management issues, and several application cases are introduced. It emphasizes on key tools and network management architecture that can support the future integration of ML into optical networks.  

This paper focuses on the applications of ML in optical network domain. We survey the papers from two major directions include optical network control and management, and optical networks status monitoring. Our contributions in this paper can be summarized as follows:

\begin{itemize}
\item \textbf{Paradigms and motivations of applying ML in optical networks:} We summarize and discuss the paradigms of applying ML in optical networks, which can be classified into three categories: regression, classification, and decision-making. The system architecture that can aid the application of ML is also discussed. Besides, we analyze the motivations and driven factors of applying ML to solve problems in optical networks from the aspects of inherent characteristics of ML algorithms, and emerging enabling techniques in the optical network domain.
 
\item \textbf{ML applications for intelligent optical networks:} We review a wide range of works, in which optical network operation and monitoring tasks are solved or aided by ML techniques. Firstly, network control and resource management related to service provision and resource assignment are reviewed. These tasks include traffic prediction and resource management. Traffic prediction helps network to act in a proactive way; and resource management will include routing assignment, Routing and Wavelength Assignment (RWA) in WDM, and Routing and Spectrum Assignment (RSA) in Elastic Optical Network (EON). Additionally, to support the functions of service provision resource assignment, intelligent monitoring of network performance is also aided by ML techniques. The intelligent monitoring includes the performance monitoring of physical layer link and signal, estimation of Quality of Transmission (QoT) of lightpath, and failure management in optical networks. These two directions are tightly connected, because the network monitoring can provide network states and performance to the control and resource assignment for better decisions.\\
\\

\item \textbf{Future challenges and possible solutions:} Since the field of applying ML in optical networks is still far from maturity and will remain being investigated, we highlight several challenges which emerge from ML applications in optical networks. These challenges refer to data open access issues, current ML model drawbacks, system security, and the reality gap between simulation and real networks. Some possible solutions are also presented, which is expected to inspire researchers with new directions.
\end{itemize}

The remainder of this paper is organized as follows. Section 2 reviews the paradigms and motivations of applying ML for intelligent optical networks. In Section 3, we give a brief overview of ML techniques used for intelligent optical networks from three aspects: supervised learning, unsupervised learning, and reinforcement learning. Then, the researches of utilizing ML in optical networks control and resource management are discussed in Section 4, whilst Section 5 reviews how the ML techniques are used in the optical network monitoring and survivability tasks to support the network control. In Section 6, the challenges of applying ML and the possible solutions in optical networks are discussed. Finally, Section 7 concludes the paper. To explain it more explicitly, the structure of this survey is depicted in Fig. 1.
\begin{figure}[htbp]
\centering
\includegraphics[width=15cm]{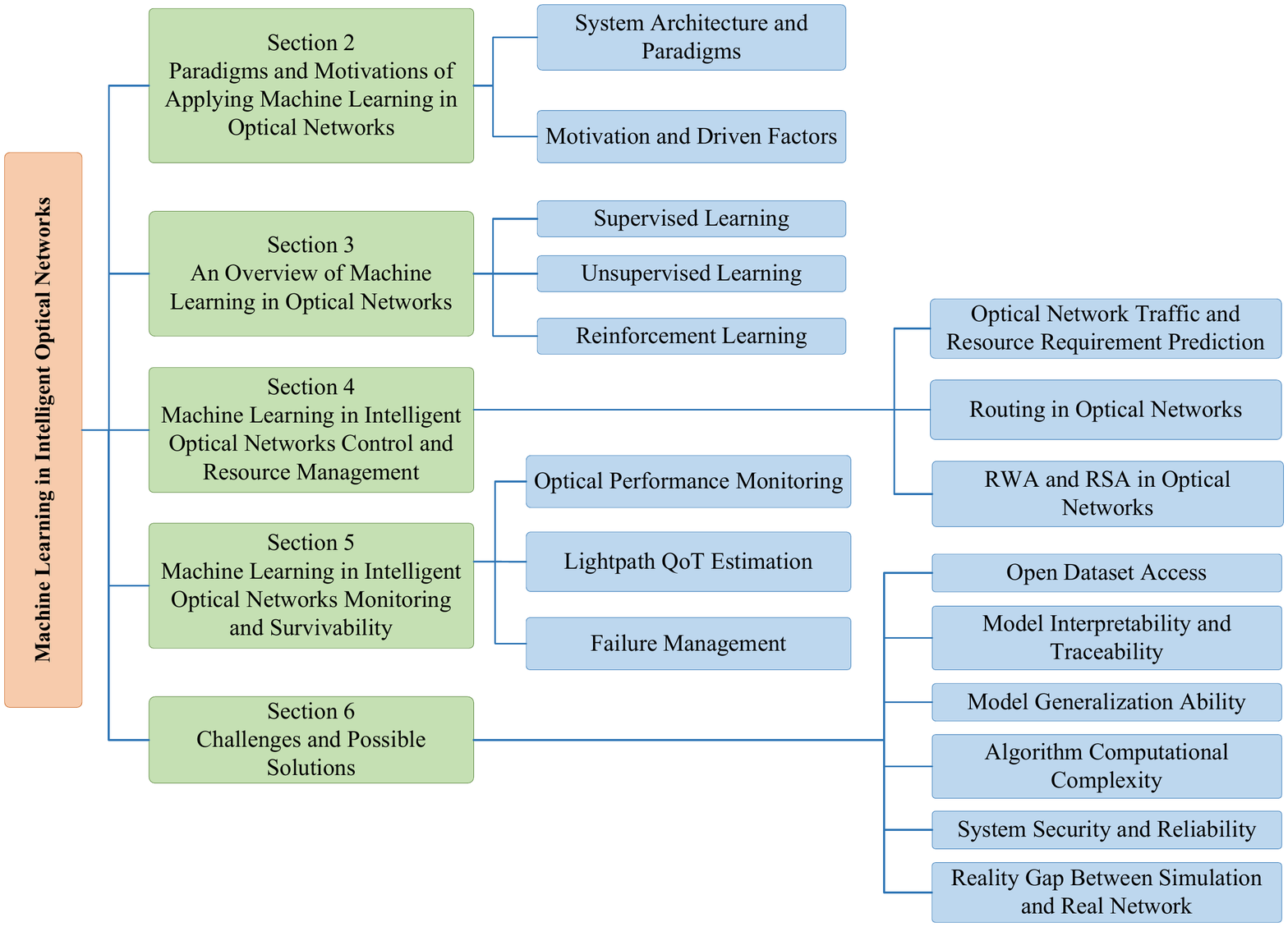} 
\caption{Structure of the survey.}
\end{figure}

\begin{figure}[htbp]
	\centering
	\includegraphics[width=15cm]{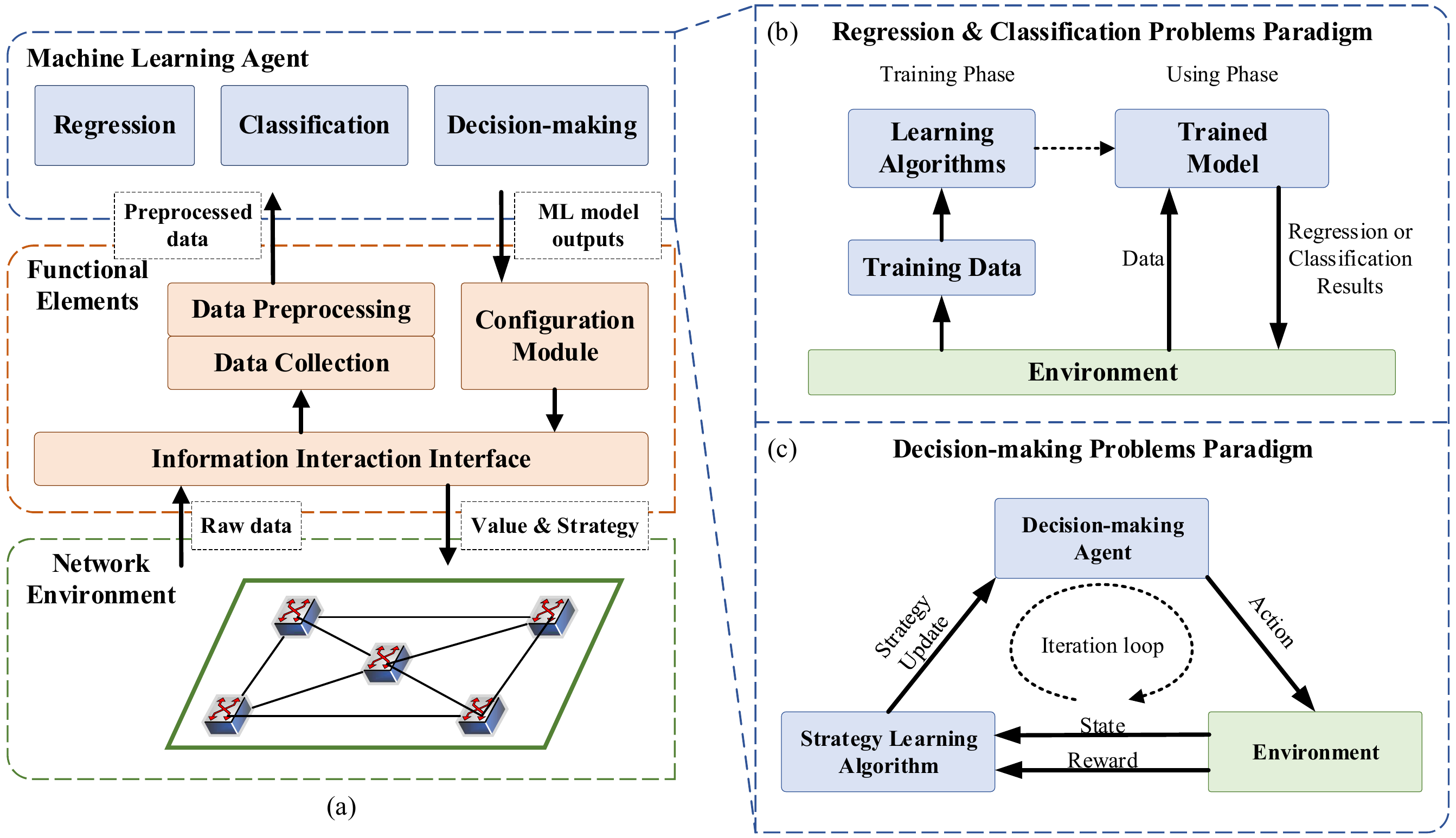} 
	\caption{(a) System architecture that incorporate with ML. (b) The learning paradigm of regression and classification problems. (c) The learning paradigm of decision-making problems.}
\end{figure}
\section{Paradigms and Motivations of Applying Machine Learning in Optical Networks}
This section summarizes the paradigms of applying machine learning in solving optical networks problems. Besides, the motivations of applying ML are also analyzed in detail, from aspects of inherent ML characteristics and the optical network changes that can support ML applications.

\subsection{System Architecture and Application Paradigms}
The system architecture of optical networks incorporating ML is depicted in Fig. 2 (a). To employ the ML-based methods in optical networks, an intelligent module that consists of Functional Elements (FEs) and ML agents should be deployed. 

The FEs are responsible for information interaction between ML agent and physical optical networks. In FEs, Data Collection module collects raw data from the optical network, and Data Processing module preprocesses it to a certain data structure used by ML models. Network protocols and functions, should be modified to support FE on network data collection \cite{JNCA-DATA-COLLECTION} and data processing.

The collected and preprocessed network data as well as network state information from FEs will be sent to ML agents for training. An ML agent may work in mainly three paradigms in optical networks: \emph{i)} regression, \emph{ii)} classification, and \emph{iii)} decision-making. The workflows of these three paradigms are depicted in Fig. 2. 

Regression and classification problems are usually solved with supervised learning and unsupervised learning. As shown in Fig. 2 (b), the ML model parameters are determined and iteratively optimized through learning algorithms with training datasets. The learning algorithms are methods that improve the model performance under a certain metric, such as Sequential Minimal Optimization (SMO) for Support Vector Machine (SVM) and backpropagation for neural networks \cite{PRML}. The well-trained ML models are used in optical networks environments, with features as input, and output the regression and classification results.

In decision-making tasks, the ML models learn the optimal strategy through interacting with the environment (optical network), as is shown in Fig. 2 (c). When an action is employed in the optical network environment, a reward is returned to the learning algorithm according to the performance of the action, and the strategies that a decision-making agent produces will be updated based on the reward. There is an iteration loop for optimal strategy learning. The learned optimal strategy decides which action to choose under a given specific state. Reinforcement learning is usually employed in decision-making problems. 

Among optical networks tasks, resource allocation is usually modeled as decision-making or action selection problems, the network prediction and monitoring are usually modeled as regression and classification problems. The important information of the works using ML techniques are summarized in Table 1, Table 2, and Table 3. The detailed discussion of specific work will be presented in section 4 and section 5. The usages of ML techniques in different optical networks tasks are summarized in Table 4.

\subsection{Motivation and Driven Factors}
There are several motivations and driven factors to support the application of ML in optical networks, as follows.
\subsubsection{Historical Data Utilization}
In recent years, various kinds of statistic data on optical network management and monitoring is accumulated. Then, how to fully use these historical data for optimizing network running become an important and emerging requirement. The traditional methods, such as Bayesian estimation methods and heuristic-based methods, usually lack of historical information usage, and only exploit current network state for optical network tasks. Also, these methods will face obvious performance degradation when there is a tiny noise or error in the samples being used.  

When employing ML, e.g. in regression and classification tasks, a dataset that contains historical information is fed to the model, and the model learns inner relationship in the dataset.  Then, the trained ML model contains the inner dependence and knowledge of historical data. As a result, ML-based method will be more robust and achieve more accurate performance against the data noise, and do not need to run the algorithm again when the network state changes in a slight range. 
\subsubsection{Reduce the Online Computation Requirements}
Facing 5G and the related services, there is a serious challenge: on one hand, the emerging services (such as uRLLC) usually require low latency; and on the other hand, the dynamic nature of the traffic ask the network to be re-configured in real time accordingly.  The conventional approaches usually need a lot of computation, which is not suitable for online network adjustment or reconfiguration. However, some ML techniques involve two stages, off-line and online. The massive model training computation can be implemented offline in the data center where there are adequate computing resources, and the online computation can be executed in the network. 

In ML techniques, the modeling training and using phases are decoupled. Thus, it can make full use of this imbalance of computational resources distribution, and meet the timeliness acquirement even with the limited resources at intermediate nodes. Although the training process of ML models may be computational expensive, the training process is offline and can be trained in datacenters with enough computational resources. And the time consumed by trained model to give results online is relative short. For example, when monitoring OSNR with Neural Networks (NNs), the NNs should be trained offline which requires huge computation resources. But the trained NN model is much faster than computing the Split Step Fourier. Thus, ML techniques have the inherent advantages of matching the computational resources distribution.

\subsubsection{Reduce Feature Engineering and Expert Knowledge Requirement}
For some optical network tasks, analytic methods have not been fully studied, and thus, only an approximate optical result will be given based on expert knowledge.

However, with the utilization of Deep Learning \cite{DL-NATURE}, which can automatically extract features from the origin data, feature-engineering step can be simplified. This can leave out requirements for professional domain knowledge and huge manual cost in feature engineering. Take OSNR monitoring with eye diagram as example, there is no explicit relationship between the pixels in eye diagrams and OSNR value, so that the ability of analytic method is limited. However, with DL \cite{CNNOSNREYE}, the raw data can be directly input into the neural network and the feature can be automatically extracted from the raw data. The accuracy of DL-based model can meet the requirements.

\subsubsection{Related Enabling Techniques}
In addition to the inherent advantages of ML techniques, there are existed enabling techniques, from aspects of network architecture, network management, and optical devices, that also provide convenient of using ML in optical networks.
From the aspect of network architecture, Software Defined Optical Network (SDON) introduces the abilities of fully programmable and reconfigurable into optical networks which increase the operational flexibility \cite{JNCA-SDN}. Besides, an architecture with cognitive plane above the control plane to support intelligent controlling is proposed \cite{CHRON}. In such novel network architectures, a ML-based function, for example a trained neural network model for OPM, can be programmed as a module in the control plane, and thus is easy to be deployed, reconfigured, and updated \cite{OPM-USE}.

From the aspect of network management and data collection, with the aid of SDN/SDON network architecture, the model-driven streaming telemetry is employed in optical network operations. It has the capability for vendor-agnostic network monitoring with gRPC protocol and YANG model \cite{TELE}. Besides, the monitored data and network configuration are transmitted through northbound and southbound in SDON; and then, a management loop in the optical network is formed which is important for ML applications.

From the aspect of optical devices, both DSP techniques and all‑optical signal processing techniques \cite{JI_ALL_OPTICAL} are becoming more mature, which makes the awareness of optical networks more efficient and accurate. Optical Time Domain Reflectometry (OTDR) has been embedded in optical devices \cite{OTDR}, which also increases the perception of network operators of optical networks. The information about the network and signal state is more convenient to be obtained and is fed to ML models for training and rewards. 

\section{An Overview of Machine Learning in Optical Networks}
ML is a hot topic in recent years, and have already been used in image recognition \cite{IR}, speech recognition \cite{SPEECH}, natural language processing \cite{NLP}, game control \cite{GO}, and recommendation systems \cite{RS}. Besides, as is described in Section 1, several communication systems are also empowered by ML. 

In this section, we give a brief introduction of ML algorithms that have been used in optical networks. The ML algorithms are classified into three categories: supervised learning, unsupervised learning, and reinforcement learning. 
\subsection{Supervised Learning}
In supervised learning, samples that consist of input vector and traget output values are input into ML models to infer a function mapping the inputs and outputs \cite{FML}. In the following, several supervised learning algorithms that are most used in optical networks will be introduced.
\subsubsection{Support Vector Machine (SVM)}
The SVM algorithm is mostly used for classification. In SVM, the training set includes $m$ samples, with each sample has the form of $({{\mathbf{x}}_{i}},{{y}_{i}})$, with $\mathbf{x}\in {{\mathbf{R}}^{N}}$ and $y\in \{-1,1\}$. The data can be separated by a hyperplane, as is shown in Fig. 3 (a), with the form of:
\begin{equation}
{{\mathbf{w}}^{T}}\mathbf{x}+b=0,
\end{equation}
where $\mathbf{w}$ is the weight vector, and $b$ is a scalar parameter \cite{FML}.
\begin{figure}[htbp]
\centering
\includegraphics[width=10cm]{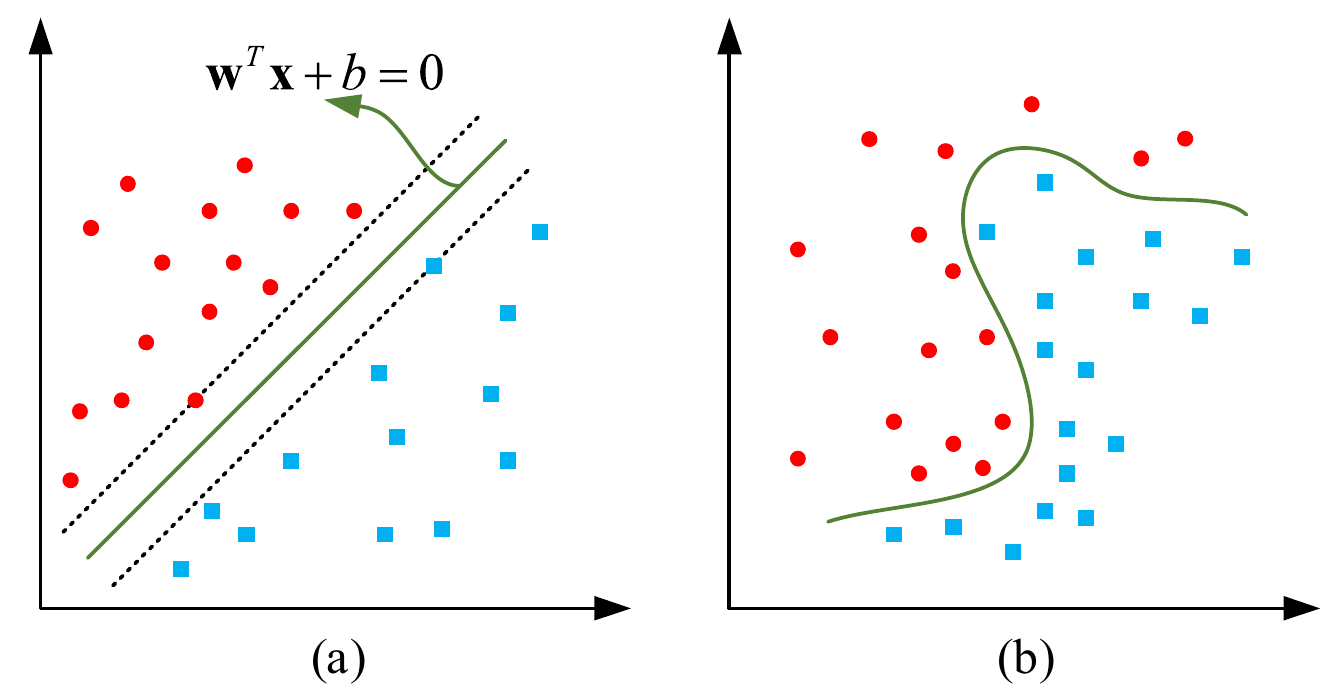}
\caption{The schematic diagram of (a) SVM with linear decision hyper-plane , and (b) nonlinear decision hyper-plane with kernel function.}
\end{figure}

The algorithm aims to find a maximal margin hyperplane. The \emph{maximal margin} means that, the hyperplane can separate the data into two classes. Meanwhile, the distance between the closest vectors to the hyperplane is maximal. The task of maximizing the margin is equivalent to minimizing the norm ${{\left\| \mathbf{w} \right\|}^{2}}$ \cite{FML}; thus, the optimization problem can be formulated as:
\begin{equation}
\begin{aligned}
& \min \; \frac{1}{2}{{\left\| \mathbf{w} \right\|}^{2}} \\
& s.t. \; \text{ }{{y}_{i}}({{\mathbf{w}}^{T}}{{\mathbf{x}}_{i}}+b)\ge 1
\end{aligned}.
\end{equation}
It is a convex quadratic programming with linear constraint. Lagrange multipliers can be introduced and the Lagrangian of the above optimization problem can be written as
\begin{equation}
L=\frac{1}{2}{{\left\| \mathbf{w} \right\|}^{2}}+\sum\limits_{i=1}^{m}{{{\alpha }_{i}}\left[ 1-{{y}_{i}}\left( {{\mathbf{w}}^{T}}{{\mathbf{x}}_{i}}+b \right) \right]},
\end{equation}
where ${{\alpha }_{k}}$ is the Lagrange multipliers \cite{FML}. And then, the dual problem can be described as
\begin{equation}
\begin{aligned}
& \begin{array} {r@{\quad}l}
{\max}_{\left( {{\alpha }_{1}},\cdots ,{{\alpha }_{m}} \right)} & \sum\limits_{k=1}^{m}{{{\alpha }_{i}}}-\frac{1}{2}\sum\limits_{i=1}^{m}{\sum\limits_{j=1}^{m}{{{\alpha }_{i}}{{\alpha }_{j}}{{y}_{i}}{{y}_{j}}\mathbf{x}_{i}^{T}{{\mathbf{x}}_{j}}}} \\
s.t.&  \sum\limits_{i=1}^{m}{{{\alpha }_{i}}{{y}_{i}}}=0 \\
&{{\alpha }_{i}}\ge0, \forall i=1,\cdots ,m \\
\end{array}
\end{aligned}.
\end{equation}
This convex quadratic programming can be solved with Sequential Minimal Optimization (SMO) efficiently \cite{PRML}. After determining the parameters, the decision function is computed as:

\begin{equation}
f\left( \mathbf{x} \right)= \rm{sgn} \left\{ {{\mathbf{w}}^{T}}\mathbf{x}+b \right\}=\rm{sgn} \left\{ \sum\limits_{i=1}^{m}{{{\alpha }_{i}}{{y}_{i}}\mathbf{x}_{i}^{T}\mathbf{x}+b} \right\}.
\end{equation}

When the data cannot be linearly separated, kernel methods can be adopted in SVM to map the origin features from a low dimension to a higher dimension space and then, the data can be linearly separated in the higher dimension space. The hyperplane may not be linear in original space, as is shown in Fig. 3 (b). Gaussian Kernel is widely used as kernel in SVM \cite{PRML}. Besides, when the data is too complex to use kernel method to generate nonlinear hyperplane, the $\delta$-margin separating hyperplanes can be introduced which allow mis-classification at some samples and guarantee better generalization ability on the whole dataset. 
\subsubsection{Neural Networks (NNs)}
NNs are effective tools when utilized in solving complex real-world problems. Its most beneficial properties include remarkable information processing ability, high parallelism, fault and noise tolerance, and strong generalization \cite{ANNFCA}. However, the disadvantages of NNs, such as sensitive to training hyper-parameters, overfitting and large computing resource requirement, should also be concerned.  NNs can be divided into three types: ANNs, CNNs, and RNNs.

\emph{a) Artificial Neural Networks (ANNs)}

An example of ANN consists of an input layer, one hidden layer, and an output layer, is shown in Fig. 4. ANNs which have more than one hidden layer are also called Deep Neural Networks (DNNs). Hidden layer and output layer are composed of neurons which receive the input vector and computes the output value through a nonlinear activation function, respectively.
\begin{figure}[htbp]
\centering
\includegraphics[width=8cm]{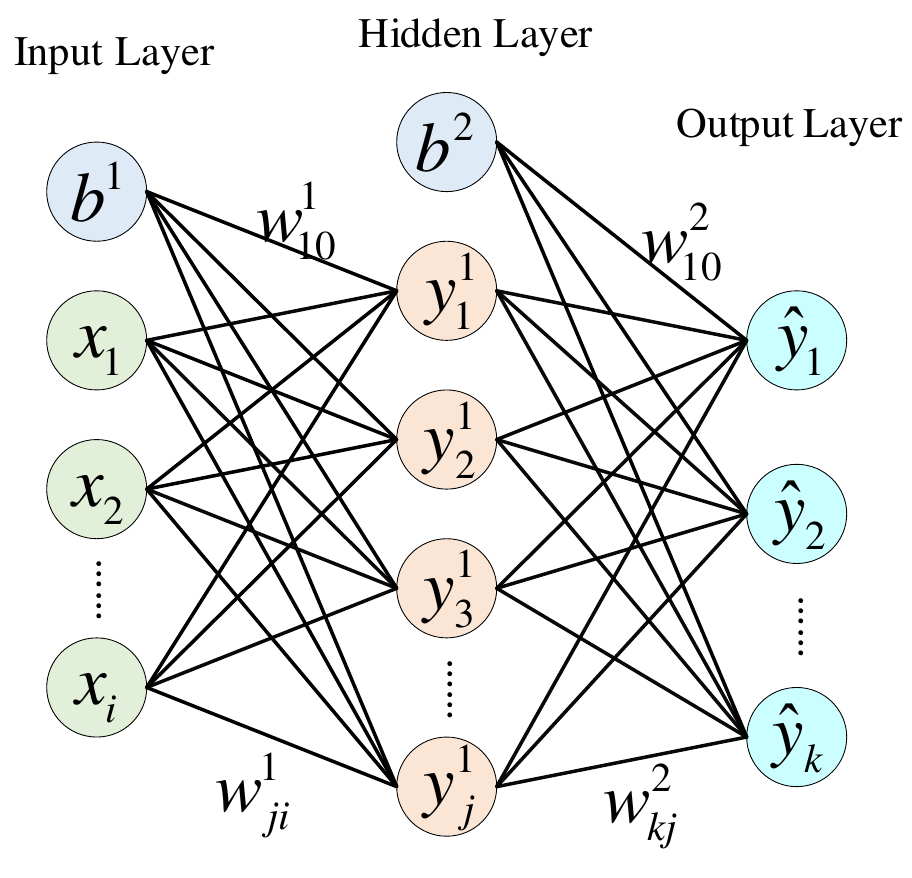}
\caption{The architecture of neural networks with one hidden layer.}
\end{figure}

As shown in Fig. 4, $y_{j}^{l}$ is the output of the ${{j}^{th}}$ neuron in the ${{l}^{th}}$ layer, $w_{ji}^{l}$ is the weight between the ${{j}^{th}}$ neuron in the ${{l}^{th}}$ layer and the ${{i}^{th}}$ neuron in the ${{\left( l-1 \right)}^{th}}$ layer, ${{b}^{l}}$ is the bias of ${{l}^{th}}$ layer, and $\mathbf{\hat{y}}$ is the final output vector \cite{NSLT}. When adopting activation function in the format of sigmoid function as 
\begin{equation}
g\left( a \right)=\frac{1}{1+{{e}^{-a}}},
\end{equation}
the output $y_{j}^{l}$ is given by 
\begin{equation}
y_{j}^{l}=\frac{1}{1+{{e}^{-\sum\nolimits_{i}{w_{ji}^{l}y_{i}^{l-1}+{{b}^{l}}}}}}.
\end{equation}
In the training process, the training performance should be evaluated by a loss function, such as the Mean Square Error (MSE):
\begin{equation}
{\rm{MSE}} {=}\frac{1}{2}\sum\limits_{k=1}^{K}{{{\left( {{{\mathbf{\hat{y}}}}_{k}}-{{\mathbf{y}}_{k}} \right)}^{2}}},
\end{equation}
where $K$ is the number of output neurons, and $\hat{y}$ and ${{y}_{k}}$ are output and target output of the sample $k$, respectively. In general, the error backpropagation mechanism is used to reduce the MSE, and improve the training performance. The update rule of weight $w_{ji}^{l}$ is as:
\begin{equation}
w_{ji}^{l}  \text{=}  w_{ji}^{l}+\eta \Delta w_{ji}^{l},
\end{equation}
where $\Delta w_{ji}^{l}=\delta_{j}^{l}y_{i}^{l-1}$, $\eta$ is the learning rate and $\delta _{j}^{l}$ is defined as
\begin{equation}
\begin{cases}
{{o}_{j}}\left( {{t}_{j}}-{{o}_{j}} \right)\left( 1-{{o}_{j}} \right),    &    \text{if layer }l\text{ is the output layer} \\
y_{j}^{l}\left( 1-y_{j}^{l} \right)\sum\nolimits_{k}{\delta _{k}^{l+1}w_{kj}^{l+1}},    &    \text{if layer }l\text{ is the hidden layer}\\
\end{cases}.
\end{equation}
It involves gradient descent to update $\mathbf{w}$ and $b$ in each updating iteration, and the iterative procedure will stop until the value of MSE is smaller than a certain threshold.

\emph{b) Convolutional Neural Networks (CNN)}

Convolutional Neural Networks (CNNs) are invented to fully use the relative fixed pattern of local characteristics. They are specialized kind of neural networks that deal with grid-like topology \cite{DL-NATURE}. Three main approaches, including \emph{spares interactions}, \emph{parameter sharing}, and \emph{equivariant representations}, are leveraged in CNNs to reduce complexity and improve performance \cite{DL}. 

A CNN consists of convolution layers, pooling layers, and fully connected layers. Each convolution layer consists a set of kernels, and each kernel with a small receptive field is used to detect local features. As is depicted in Fig. 5 (a), “Convolution” in CNNs indicates each kernel convolves with pixel points across the width and height of input image, and computes the summation of dot product between the entries of the kernel and the input. In pooling layer, the outputs of prior layer are replaced with a summary statistic of nearby outputs. The max pooling, which is most used in CNNs, is shown in Fig. 5 (b). The max pooling kernel output the maximum value within the corresponding rectangular area of image. At the end of CNNs, fully connected layers are used to compute the final classification or regression results.
\begin{figure}[htbp]
\centering
\includegraphics[width=11cm]{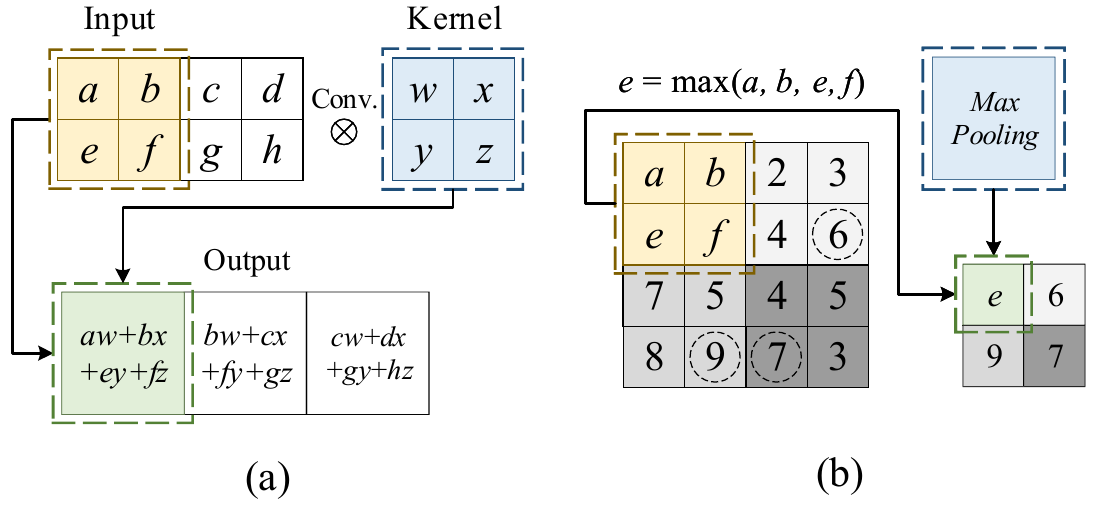}
\caption{A schematic diagram of (a) Convolutional kernel, and (b) Pooling kernel with max pooling.}
\end{figure}

Training frameworks, such as TensorFlow \cite{TF}, Caffe \cite{CAFFE}, and Keras \cite{KERAS}, are powerful tools for setting up network architecture and CNN training. Besides, there have existed mature CNN architectures such as LeNet \cite{LENET}, VGG \cite{VGG}, and ResNet \cite{RESNET}, which have achieved great performance in image processing. Note that, these successful models can be used in optical networks tasks with fine tune procedure and then, a repetition of CNN network architecture design may be left out.

\emph{c) Recurrent Neural Networks (RNN)}

To capture characteristics and dependence in sequence-form data, Recurrent Neural Networks (RNNs) are invented. RNNs are used for processing sequential data, such as audios, sentences, and texts, which are rich in contextual information. RNNs can map a sequence data to another sequence data with the uniform or non-uniform length \cite{DL}. As is depicted in Fig. 6, in RNN, parameter sharing is realized by using previous outputs as the inputs. Therefore, the output of RNN at a certain moment depend on all previous inputs \cite{DL}.

The most popular transformation of RNN is Long Short-Term Memory (LSTM) which introduces gated memory and internal self-loops to its hidden units. With such a granular internal processing unit, LSTM stores and update the contextual information efficiently. LSTM can overcome the weakness of conventional RNNs, such as the backpropagation gradient vanishing and blow-up issues \cite{DL}. 

More improvement works in LSTM include Bi-directional LSTM (BLSTM) \cite{BLSTM}, Gated Recurrent Unit (GRU) \cite{GRU}, CNN combined with LSTM \cite{CLSTM}, and LSTM with Attention mechanism \cite{ATTENTION}. 
\begin{figure}[htbp]
\centering
\includegraphics[width=10cm]{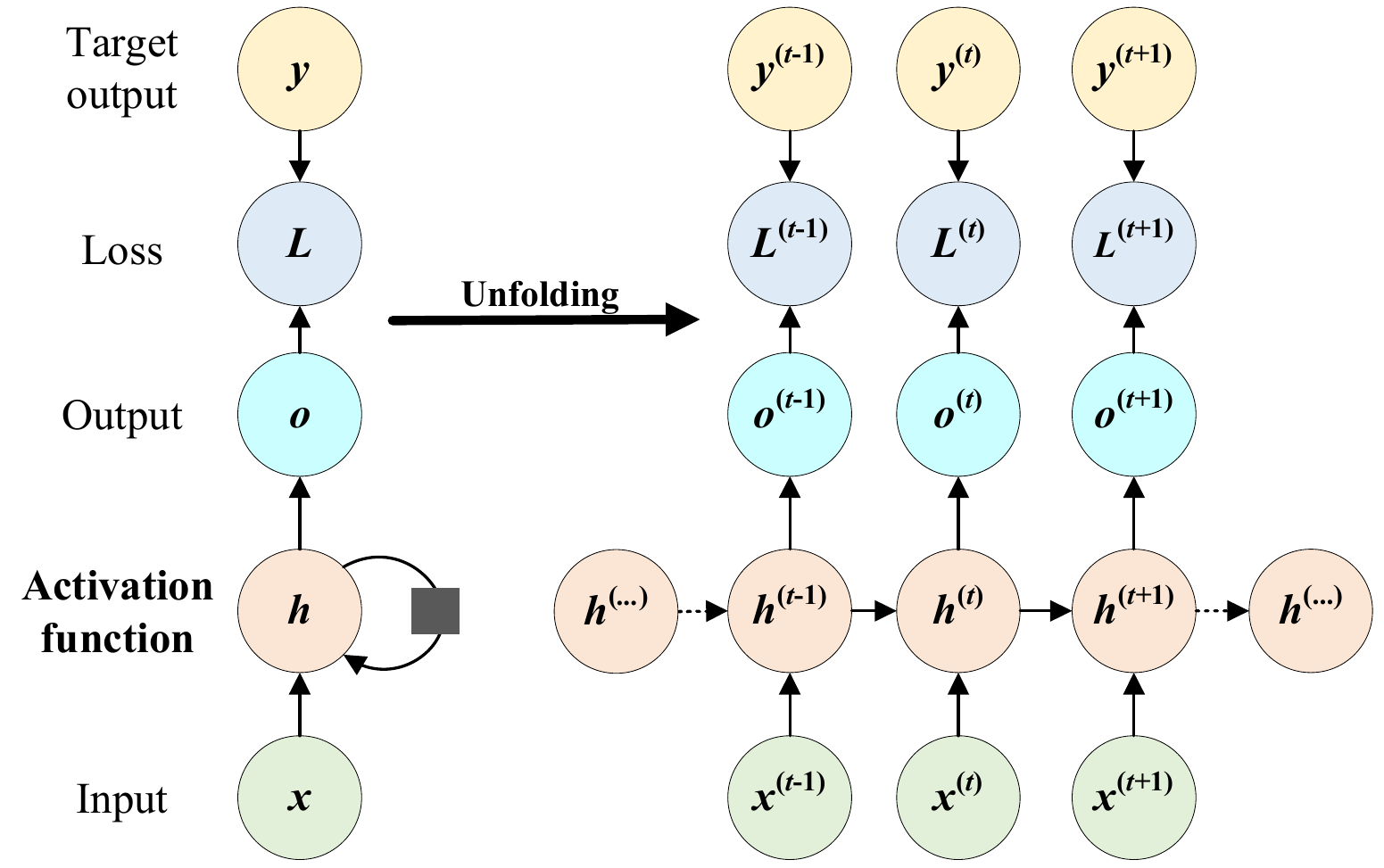}
\caption{The architecture of RNN.}
\end{figure}
\subsection{Unsupervised Learning}
Unsupervised learning uses the unlabeled dataset for training. Basically, the goals of unsupervised learning include \cite{PRML}:

\begin{itemize}
\item {Clustering:} Classifying data into different groups by according to the similarity among them, such as K-means.
\item Dimensionality reduction: Projecting a high-dimensional data down to a low-dimensional space, such as Principal Component Analysis (PCA).
\item Density and parameter estimation: Determining the distribution of data, or estimating unknown parameters in distribution, such as Expectation Maximization (EM).
\end{itemize}

In the following sub-sections, we will discuss K-means, and PCA, which both have been used in optical networks. EM has been used in optical communications, such as signal detection \cite{EM-MLTOC}, and signal nonlinear compensation \cite{EM-NONLINEAR}, but to the best of our knowledge, it has not been used in optical networks yet.
\subsubsection{K-means Clustering}
K-means clustering is widely used for classification problems with unlabeled data. The ``$K$'' here refers to the number of clusters. A dataset in K-means clustering including $m$ samples has the form of $\{{{\mathbf{x}}_{i}}\}_{i=1}^{m}$ with $\mathbf{x}\in {{\mathbf{R}}^{N}}$. A set of N-dimensional vectors ${{\mathbf{\mu }}_{k}}$ is first introduced, where $k=1,\cdots ,K$, is the prototype associated with the ${{k}^{th}}$ cluster. The goal of K-means is to minimize the inner-class distance $J$:
\begin{equation}
J=\sum\limits_{k=1}^{K}{\sum\limits_{i=1}^{m}{{{r}_{ik}}}}{{\left\| {{\mathbf{x}}_{i}}-{{\mathbf{\mu }}_{k}} \right\|}^{2}},
\end{equation}
where ${{r}_{ik}}\in \left\{ 0,1 \right\}$ indicates whether ${{\mathbf{x}}_{i}}$ belongs to the ${{k}^{{th}}}$ cluster.
The optimal procedures are as follows \cite{PRML}:

\emph{i)} Initialize \emph{K} cluster prototypes, in the form of ${{\mathbf{\mu}}_{k}}\in {{\mathbf{R}}^{N}},\text{ }k=1,\ldots ,K$.

\emph{ii)} Assign each sample in the dataset to the nearest cluster. The distance indicator is Euclidean distance with the form:
\begin{equation}
Dij=\sqrt{\sum\limits_{j=1}^{N}{{{\left| {{x}_{ij}}-{{\mu }_{kj}} \right|}^{2}}}}.
\end{equation}
Thus, the assignment rule is 
\begin{equation}
r_{ik}\text{=} 
\begin{cases}
1,    &    \text{if }k\text{ = }\underset{j}{\mathop{\text{arg}}}\,\text{ min}{{\left\| \mathbf{x}_i-{{\mathbf{\mu }}_{j}} \right\|}^{2}}\\
0,    &    \text{otherwise }\\
\end{cases}.
\end{equation}

\emph{iii)} Update the cluster prototype ${{\mathbf{\mu }}_{k}}$ with 
\begin{equation}
{{\mathbf{\mu }}_{k}}=\frac{\sum\nolimits_{i}{{{r}_{ik}}{{\mathbf{x}}_{i}}}}{\sum\nolimits_{i}{{{r}_{ik}}}}.
\end{equation}

\emph{iv)} Repeat step \emph{ii)} and step \emph{iii)} until there is no change in the assignments (or the maximum number of iterations is reached).

Although K-means clustering performs perfectly in practice, there are still some issues to be concerned when using it: \emph{i)} There is no efficient method to pre-identify the number of clusters, and usually   is predefined according to the problems. \emph{ii)}  The optimization procedure cannot guarantee converge to a global optimum. In practical applications, algorithms should be run several times with different starting point to get the optimal solution. \emph{iii)} K-means clustering is sensitive to noise and abnormal data \cite{SCA}.
\subsubsection{Principal Component Analysis (PCA)}
PCA is usually used for data compression, dimensionality reduction, and feature extraction. PCA always plays a role as data pre-processing to reduce the complexity of subsequent tasks. The goal of PCA is projecting the original N-dimensional data onto an M-dimension space (M$<$N), with least information loss \cite{PRML}. 

The original data in PCA has the form of ${{\mathbf{x}}_{i}}$ with $\mathbf{x}\in {{\mathbf{R}}^{N}}$. To guarantee that the projected samples can be easily classified, the variance of new data should be maximized. To find the first principal component, considering the projection onto a one-dimensional space (M=1). The new component ${{\xi }_{1}}$ will be the linear combination of the original feature with the formulation: 
\begin{equation}
{{\xi }_{1}}\text{ = }\sum\limits_{j=1}^{N}{{{\alpha }_{1j}}\text{ }{{x}^{j}}}\text{= }{{\mathbf{\alpha }}_{1}}^{T}\mathbf{x},
\end{equation}
where ${{\mathbf{\alpha }}_{1}}$ is an N-dimensional unit vector so that ${{\mathbf{\alpha }}_{1}}^{T}{{\mathbf{\alpha }}_{1}}=1$. The variance of ${{\xi }_{1}}$ is given by
\begin{equation}
\operatorname{var}({{\xi }_{1}})=\frac{1}{l}\sum\limits_{i=1}^{m}{{{\{{{\mathbf{\alpha }}_{1}}^{T}{{\mathbf{x}}_{i}}-{{\mathbf{\alpha }}_{1}}^{T}{{{\mathbf{\bar{x}}}}_{i}}\}}^{2}}}={{\mathbf{\alpha }}_{1}}^{T}\mathbf{\Sigma }{{\mathbf{\alpha }}_{1}},
\end{equation}
where $\mathbf{\Sigma }=\text{E}\left\{ (\mathbf{x}-{{\mathbf{\alpha }}_{i}}){{(\mathbf{x}-{{\mathbf{\alpha }}_{i}})}^{T}} \right\}$ is the data covariance matrix. The maximization of project variance ${{\mathbf{\alpha }}_{1}}^{T}\mathbf{\Sigma }{{\mathbf{\alpha }}_{1}}$ under constraint ${{\mathbf{\alpha }}_{1}}^{T}{{\mathbf{\alpha }}_{1}}=1$ is as
\begin{equation}
\begin{aligned}
& \begin{array} {r@{\quad}l}
{\max}  &  {{\mathbf{\alpha }}_{1}}^{T}\mathbf{\Sigma }{{\mathbf{\alpha }}_{1}}\\
  s.t.  &  {{\mathbf{\alpha }}_{1}}^{T}{{\mathbf{\alpha }}_{1}}=1\\
\end{array}
\end{aligned}
\end{equation}
The Lagrange multiplier ${{\lambda }_{1}}$ can be introduced and the goal is
\begin{equation}
\max \; {{\mathbf{\alpha }}_{1}}^{T}\mathbf{\Sigma }{{\mathbf{\alpha }}_{1}}+{{\lambda }_{1}}\left(1-{{\mathbf{\alpha }}_{1}}^{T}{{\mathbf{\alpha }}_{1}}\right).
\end{equation}
By setting the derivative with respect to ${{\mathbf{\alpha }}_{1}}$ equals to zero, the maximization can be achieved when
\begin{equation}
\mathbf{\Sigma }{{\mathbf{\alpha }}_{1}}={{\lambda }_{1}}{{\mathbf{\alpha }}_{1}}.
\end{equation}
This formulation indicates that $\mathbf{\alpha }_1$ and $\lambda_1$ is an eigenvector and an eigenvalue of the $\mathbf{\Sigma }$ , respectively. Then, the maximized variance is given by
\begin{equation}
{{\mathbf{\alpha }}_{1}}^{T}\mathbf{\Sigma }{{\mathbf{\alpha }}_{1}}={{\lambda }_{1}}
\end{equation}
The variance will be a maximum when setting ${{\mathbf{\alpha }}_{1}}$ as the eigenvector having the largest eigenvalue ${{\lambda }_{1}}$.

The procedures above find the most important principal component which is one-dimensional. The general case of an M-dimensional projection space can be defined by the \emph{M} eigenvectors ${{\mathbf{\alpha }}_{1}},\ldots ,{{\mathbf{\alpha }}_{M}}$ of $\mathbf{\Sigma }$, corresponding to the \emph{M} largest eigenvalues ${{\lambda }_{1}},\ldots ,{{\lambda }_{M}}$.
\subsection{Reinforcement Learning}
RL is widely used in control systems to optimize the decisions. It is a ``rial-and-error'' approach that the learning agents learn optimal decisions by interacting with the environment. The ``trial-and-error'' rule means RL agents make a trade-off between known decision exploitation and new decision exploration to achieve optimal policy.

As is shown in Fig. 7 (a), a RL problem is characterized by the following parameters \cite{RL}: 
\begin{itemize}
\item \emph{T}: The set of iteration times. \emph{T} includes a sequence of discrete time steps. At each time step \emph{t}, the agent completes an iteration with the environment.
\item \emph{S}: The finite set \emph{S} includes the possible states of the environment.
\item \emph{A}: The finite set $A\left( {{s}_{t}} \right)$ is the set of actions available in state ${{s}_{t}}$.
\item ${{p}_{t}}\left( {{s}_{t+1}}|{{s}_{t}},{{a}_{t}} \right)$ denotes the state transition probability that environment transfer from ${{s}_{t}}$ to ${{s}_{t\text{+}1}}$ under action ${{a}_{t}}\in A\left( {{s}_{t}} \right)$.
\item $\pi$ is the policy that map from \emph{S} and action \emph{A}. $\pi \left( s,a \right)$ denotes the probability of acting $a$ under state $s$.
\end{itemize}

In the ${{t}^{th}}$ iteration, the agent observes the current environment state ${{s}_{t}}$, and chooses an action ${{a}_{t}}$. After that, the environment transfers from the state ${{s}_{t}}$ to ${{s}_{t\text{+}1}}$ following the probability ${{p}_{t}}\left( {{s}_{t+1}}|{{s}_{t}},{{a}_{t}} \right)$, and returns a reward ${{r}_{t}}\left( {{s}_{t}},{{a}_{t}} \right)$ according to the performance of ${{a}_{t}}$.

The objective of the iterations is finding the optimal policy ${{\pi }^{*}}$ that maximizes the expected return. But in most cases, ${{\pi }^{*}}$ is determined by seeking the maximization of expected discounted return:
\begin{equation}
R\left( t \right)=\sum\limits_{k=0}^{\infty }{{{\gamma }_{k}}{{r}_{t+k+1}}\left( {{s}_{t+k}},{{a}_{t+k}} \right)},
\end{equation}
where $\gamma \in \left[ 0,1 \right)$ is the discount rate.

The criteria to evaluate a policy $\pi $ are state-value function ${{V}^{\pi }}\left( s \right)$ and action-value function ${{Q}^{\pi }}\left( s,a \right)$. A state-value function ${{V}^{\pi }}\left( s \right)$ is the expected return when starting in s and following $\pi$, which can be defined as:
\begin{equation}
{{V}^{\pi }}\left( s \right)={{E}_{\pi }}\left\{ \sum\limits_{k=0}^{\infty }{{{\gamma }^{k}}{{r}_{t+k+}}_{1}\left( {{s}_{t+k}},{{a}_{t+k}} \right)|{{s}_{t}}=s} \right\}.
\end{equation}
Similarly, action-value function ${{Q}^{\pi }}\left( s,a \right)$ is defined as the expected value of taking action ${{a}_{t}}$ in state ${{s}_{t}}$ under the policy $\pi $, which can be described as:
\begin{equation}
{{Q}^{\pi }}\left( s,a \right)={{E}_{\pi }}\left\{ \sum\limits_{k=0}^{\infty }{{{\gamma }^{k}}{{r}_{t+k+}}_{1}\left( {{s}_{t+k}},{{a}_{t+k}} \right)|{{s}_{t}}=s,{{a}_{t}}=a} \right\}.
\end{equation}
When the optimal policy ${{\pi }^{*}}$ is obtained, the ${{V}^{\pi }}\left( s \right)$ and ${{Q}^{\pi }}\left( s,a \right)$will also be maximized as:
\begin{equation}
\begin{matrix}
   \begin{matrix}
   {{V}^{*}}\left( s \right) & = & \underset{\pi }{\mathop{\arg }}\,\max {{V}^{\pi }}\left( s \right)  \\
\end{matrix}  \\
   \begin{matrix}
   {{Q}^{*}}\left( s,a \right) & = & \underset{\pi }{\mathop{\arg }}\,\max {{Q}^{\pi }}\left( s,a \right)  \\
\end{matrix}  \\
\end{matrix}
\end{equation}

To optimize the policy $\pi$, three fundamental methods are provided: dynamic programming, Monte Carlo methods, and Temporal Difference (TD) learning \cite{RL}. Especially, Q-leaning in TD is widely used policy optimization. It does not require the knowledge of the transition probability ${{p}_{t}}\left( {{s}_{t+1}}|{{s}_{t}},{{a}_{t}} \right)$, which is always hard to know in practice. The one-step Q-learning is defined as:
\begin{equation}
Q\left( {{s}_{t}},{{a}_{t}} \right)\leftarrow \left( 1-\alpha  \right)Q\left( {{s}_{t}},{{a}_{t}} \right)+\alpha \left[ {{r}_{t+1}}\left( {{s}_{t}},{{a}_{t}} \right)+\gamma \underset{a}{\mathop{\max }}\,Q\left( {{s}_{t+1}},a \right) \right].
\end{equation}
To achieve a convergence of ${{Q}^{*}}$, all state-action pairs need to be continuously updated. This shows the exploration aspect of RL. $\varepsilon $-greedy method are adopted to seek a trade-off between exploration and exploitation. It will choose new action $a$ with probability $\varepsilon $ in each iteration even if another action a with a current optimal reward is already known. The $\varepsilon $-greedy policy selects action \emph{a} as follows:

\begin{equation}
{{a}^{*}}\left( s \right) \text{=} 
\begin{cases}
a_{c}^{*}, & \text{with Pr=1-}\varepsilon \\
\sim U\left( A\left( {{s}_{t}} \right) \right), & \text{with Pr=}\varepsilon \\
\end{cases},
\end{equation}
where $a_{c}^{*}=\arg \text{ }{{\max }_{a\in A}}Q\left( s,a \right)$  is the current optimal action, but may not be the optimal action. $U\left( A\left( {{s}_{t}} \right) \right)$ is the discrete uniform probability distribution over the action set $A\left( {{s}_{t}} \right)$, and ${{a}^{*}}\left( s \right)=\sim U\left( A\left( {{s}_{t}} \right) \right)$ means selecting the action from total action set $A\left( {{s}_{t}} \right)$ with an equal probability.
\subsubsection{Deep Reinforcement Learning (DRL)}
A practical decision scenario usually has a large state space and large action space. It is difficult for conventional RL to model such complex environment. Deep Reinforcement Learning (DRL) use the deep learning to better represent environment and actions. The architecture of DRL with Q-learning, which is a typical example of DRL, is shown in Fig. 7 (b). Deep Q-networks, such as CNN and RNN, are introduced to replace the Q-table in conventional RL to achieve better abstraction of the environment states, and learn the expectation values of each decision. The environment feeds the deep Q-network with observed state and corresponding action award for the network training \cite{DRL}.

In addition to deep Q-network, other strategies further improve the performance of DRL, including Policy Network and Monte Carlo Tree Search \cite{ALPHAGO}. Framework for DRL are provided including TensorFlow \cite{TF}, OpenAI-Baseline \cite{BASELINE} and PARL \cite{PARL}.
\begin{figure}[htbp]
\centering
\includegraphics[width=10cm]{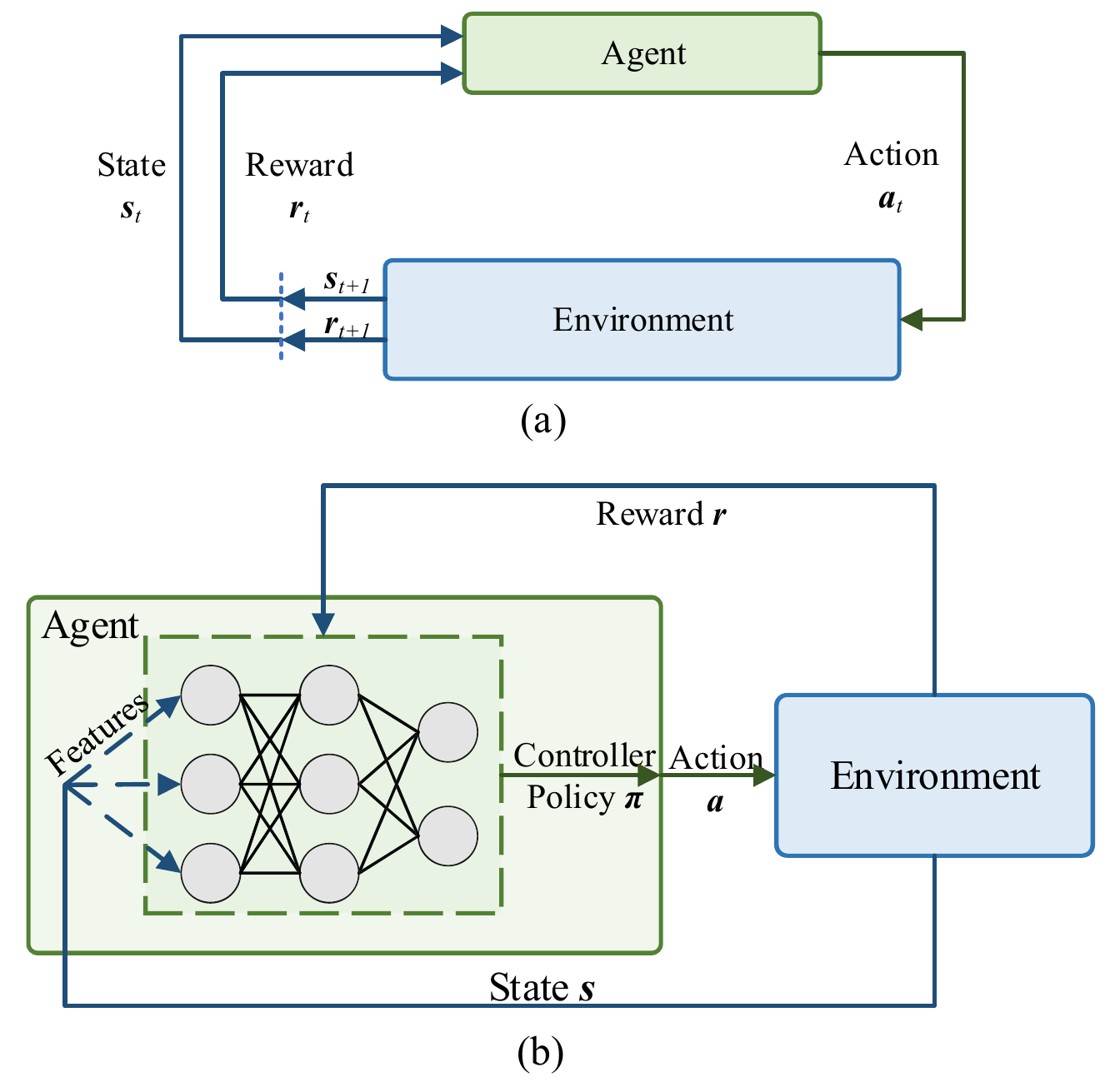}
\caption{(a) Reinforcement learning architecture. (b) Deep reinforcement learning architecture with deep neural networks replacing the traditional Q-table.}
\end{figure}

\section{Machine Learning for Intelligent Optical Networks Control and Resource Management}
Optical networks are required to receive different service requests, and map upper-layer service requests to the underlying physical resources configurations with QoS guarantee. Thus, the optical networks should consider both traffic characteristics and resource management. In this section, the ML approaches which have been used in traffic prediction and resource allocation, will be reviewed. The important information of these works refer to Table 1.
\subsection{Optical Network Traffic and Resource Requirement Prediction}
Traffic prediction can be used for network (re)configuration and resource (re)allocation in advance for future network traffic. With the predicted traffic value and future resource requirement, the network resources can be precisely allocated, and indeed the flexibility and agility of optical networks increase. 

The traffic in optical networks often presents certain regularity and periodicity, and the historical data of traffic is easy to obtain by operators. So, it is suitable use ML to fit the future traffic volume with input of traffic values in a history window.

The traffic pattern in Inter-Datacenter Optical Networks (IDCONs) that connect different datacenters, are dynamic and bursty with variations in temporal and spatial domains. Thus, the prediction of the traffic and resource requirement is essential for adaptive Network Control and Management (NC\&M). Guo \emph{et al.} propose a DNN-based method for bandwidth resource requirement prediction in IDCONs \cite{DLID}. In Virtual Optical Networks (VON), the model is trained with historical data contains traffic information, such as sources node, destinations node, and the bandwidth requirement prediction in next time slot. The predicted traffic resource requirement values are used by Infrastructure Provider (InP) to decide whether the network resource allocations should be reconfigured. When the InP monitors a significant mismatch between current allocated resources and future traffic, the InP will launch the VON reconfiguration to prepare for future traffic. Transfer learning is also used for DNN model modification and performance maintenance with small time complexity increase. The performance of the proposed method shows better performances in blocking probability and spectrum utilization, compared to fixed capacity allocation method. Besides, the authors investigate vulnerabilities of DNN training with ``Machine-Learning-as-a-Service (MLaaS)''. A data poisoning scheme is proposed to demonstrate that, the DNN may be contaminate with adversarial samples and degrade the performance of resource reallocation.

ML-based traffic prediction methods are also employed in intra-datacenters optical networks. Yu \emph{et al.} present a DNN-based traffic prediction method in datacenter networks, and use the predicted traffic information for resource allocation to improve the resource utilization of high bandwidth and decrease the blocking probability \cite{LDLA}. The predicted traffic properties include traffic arrival time and resource consumption. With these predicted properties, the traffic queue can be adjusted to achieve lower path blocking probability.

To efficiently transport different granularity of traffic flows with low congestion, the hybrid electro-optical DCN architectures are proposed. The hybrid electro-optical DCN consists of Optical Circuit Switching (OCS) to transfer heavy traffic, and Electrical Packet Switching (EPS) to transfer short-lived and burst data flows. Traffic prediction is essential to early decision of whether a flow should transmit by OCS or EPS. In \cite{HIDCN}, a Nonlinear Autoregressive Neural Network (NARNN) is used for traffic prediction in hybrid electro-optical DCN scenarios with heavy traffic. The predicted traffic information makes it possible to prioritize the future heaviest traffic streams for optical switching and offloading the EPS traffic.

The traffic data is in the form of sequence, so it is suitable to be processed with RNN models. Singh \emph{et al.} demonstrate the using of LSTM for traffic remaining time prediction, which provide information for traffic aggregation in optical datacenter networks \cite{MLBP}.When the resource is allocated or re-allocated in data centers, it is essential to know the Mean Residual Life (MRL) which is a function of the spent time and Holding Time Information (HTI). Due to the heterogeneity and diversity nature of the applications in data centers, the HTI of traffics are not known. While the tradition methods assume an exponentially distribution for HTI \cite{MLBP9}, it is not valid for data center traffic because of heavy tail characteristic. Thus, using LSTM network to learn the nonlinear relationship between historical traffic and HTI is reasonable.
\subsection{Routing in Optical Networks}
Route planning is one of the fundamental tasks in optical networks. With the scale and the complexity of optical networks increase, the conventional Shortest Path First (SPF) routing algorithm may result in low network resource utilization and high blocking ratio. Heuristic-based route planning will suffer high computational complexity when facing large scale topology. 

To overcome the drawbacks of simple SPF routing and heuristic-based routing, ML techniques have been employed. Some works model the routing allocation as classification and regression tasks, which use supervised learning to obtain the rules of routes generation  from the historical route dataset \cite{GPRM-OBS} \cite{MLAR} \cite{WITHOUT}. Other works model the routing problem as decision-making tasks, in which the RL is employed to generate optimal routing assignment \cite{NRL-LP} \cite{MARL-OBS} \cite{RLDRS} \cite{PQDR}. In this section, applications of ML for simple routing tasks are presented. Note that, a further discussion of the joint resource assignment, include the assignment of route, wavelength, spectrum, and modulation format, will be reviewed in section 4.3. 
\subsubsection{Supervised Learning-based Routing}
In supervised learning-based routing, the routing tasks are modeled as classification or regression problems. The training dataset are obtained from historical route sets or pre-computed routes with ILP.  

In \cite{GPRM-OBS}, Graphic Probabilistic Routing Model (GPRM) which is based on Bayesian Network (BN) is presented to select less utilized links in an Optical Burst Switching (OBS) network to reduce Burst Loss Ratio (BLR) without affecting the end-to-end delay.  A BN model is exploited at each node in the network, which determines the next hop according to a routing table updated by the BN. The routing table contains several tuples of input burst information vector, next hop, and its corresponding cost. The experimental results showed that the GPRM method achieves a lower burst loss ratio compared to the SPF method. 

Troia \emph{et al.} model the routing problem in SDN-based optical networks as a classification task to decide which routing plan is suitable for current traffic matrices \cite{MLAR}. Machine Learning Routing Computation (MLRC) module captures traffic matrices from networks using REST APIs, and classify the current traffic matrices into different classes, and each class corresponds to a pre-computed optimal routing solution. Net2Plan network optimization tools \cite{N2P} are employed to compute the optimal routing sets under specific traffic metrices with Integer Linear Programming (ILP). Traffic metrices are input into the logistic regression model for training and the model output the optimal routing set. The advantage of MLRC is supporting real-time network configuration. Without the computation of solving ILP, the routing decision process can be accomplished in only 80ms, from traffic matrix acquirement to installing the flow rules in SDN. 

In multi-domain network scenarios, the intra-domain network status is protected from exterior access for security and privacy considerations, and thus it is hard for centralized controllers to retrieve all network status inside domains. To obtain a route without knowing intra-domain private network information, Zhong \emph{et al.} use LSTM to learn the route generation rules from sparse historical route trajectories in multi-domain scenarios, and directly return a feasible inter-domain route \cite{WITHOUT}. The training input of LSTM are public available, such as traffic requests, historical route trajectories, and inter-domain link capabilities. The target output is the route set computed with BRPC. The deep neural networks excavate the complicated relationship between public network status and optimal routes, and will not divulge the private domain information when generating routes.
\subsubsection{Reinforcement Learning-based Routing}
When employing RL in routing planning, the tasks are modeled as decision-making problems. In decision-making problems, the learning models interact with network environment to learn the optimal actions under specific network states. RL is usually used as optimal actions learning algorithm in decision making agents.

Belbekkouche \emph{et al.} propose a Reinforcement Learning-based Alternative Routing (RLAR) model in OBS networks \cite{NRL-LP}. In this approach, all nodes (including edge nodes and intermediate nodes) are deployed with learning agents which choose the optimal next output link at each node. Each learning agent contains a lookup table called Q-table that stores the pairs (destination, neighbor) and corresponding Q-value for the next hop selection of the burst. When the bursts contended at the node, RLAR simply drops one of the contention bursts rather than deflects them. The $\varepsilon $-greedy and Q-learning are used to find the optimal policy and obtain model convergence.

In single agent system, each agent makes decisions without considering the actions other agents select, which may result in sub-optimal in entire networks. Kiran \emph{et al.} proposed a Multi-Agent Reinforcement Learning (MARL) approach for path selection in OBS, where all agents, which equipped at ingress nodes, consider actions of other nodes when making their own action \cite{MARL-OBS}. In such scenario, agents will upload their Q-tables to a central server, which will check whether the optimal route of each node shares some common links.  If the number of optimal routes which share the same link exceeds a threshold, this link may suffer heavy traffic and congestion. At this time, the second optimal route with the second largest Q-value in each node is selected for burst transmission. This method maximizes the joint utility function of all the agents in the network. Experiments results show that, the MARL has a lower burst loss probability and a higher link utilization compared with the single agent method in entire network.
\subsubsection{Reinforcement Learning-based Deflection Routing in OBS}
In OBS networks, the wavelength contention at intermediate nodes is the main cause of burst losses. There are mainly four methods for solving the wavelength contention problem: \emph{i)}  using Fiber Delay Lines (FDLs) to delay the blocked burst until there is an idle wavelength for transmission; \emph{ii)} wavelength conversion when contention exists at the intermediate node \cite{RL-DROBSN14}; \emph{iii)} burst segmentation method which segments the burst into two part with one part being dropped or forwarded on an alternate path, while the other part forwarded on the primary path \cite{SEGMENTATION}; \emph{iv)} deflection routing where only one burst is routed to its primary route while other bursts are switched to alternate routes. Due to the fact that the optical devices, such as FDLs and optical circuit, are not mature, which hinder the application of the first three methods, the deflection routing method is the hottest directions for solving wavelength contention in OBS networks \cite{DEFLECTION} \cite{DEFLECTION2002}.

Belbekkouche \emph{et al.} propose a RL-based Deflection Routing Scheme (RLDRS) to select an optimal deflection route dynamically toward a given destination \cite{RLDRS}. Each node stores a series of Q-values, which represent its appreciation of a deflection output link to a certain destination, in a Deflection Table (DT) for next hop selection when wavelength contention occurs. As is shown in Fig. 8, each entry in the DT is indexed by the pair (destination, neighbor) and when the node $x$ decides to deflect a burst, it chooses the next hop y with largest Q-value. Q-learning algorithm \cite{RL} is used to find the optimal policy as well as guarantee the convergence of the RL algorithm. RLDRS also considers the additional traffic caused by deflection routing which may degrade the network performance. To avoid extreme cases, for example, the burst is deflected for too many times, a maximum number of authorized deflections $Maxdf$ is set. When the total deflection number of a burst is larger than $Maxdf$, the burst can be dropped to prevent excessive deflections. 
\begin{figure}[htbp]
\centering
\includegraphics[width=10cm]{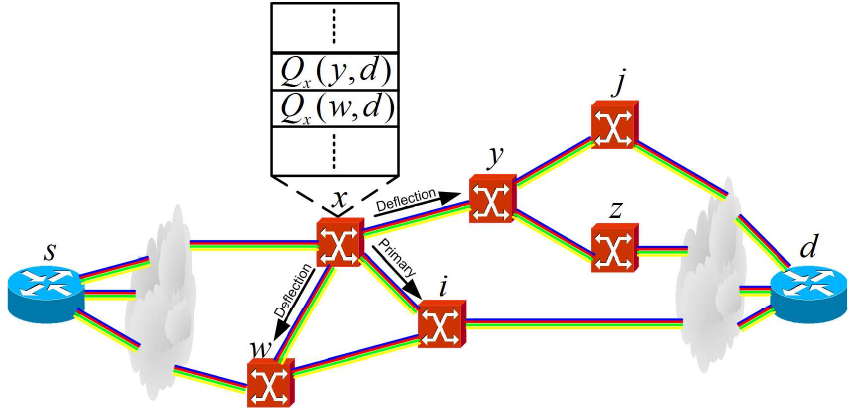}
\caption{An example of Deflection Routing and Deflection Table.}
\end{figure}

An extension of RLDRS, called IRLRCR, which integrates RLDRS with RLAR is presented in \cite{NRL-LP}. RLAR is adopted as the proactive approach to select routes, and RLDRS is adopted as the reactive approach to deflect the burst when wavelength contention occurs. Each node has only one lookup table, which is called Global Table. Global Table is used as the Q-table and deflection table in RLAR and RLDRS respectively. When a burst request arrives, the nodes use RLAR to selcet next output link with the highest Q-value in Global Table; and when the contention occurs, the RLDRS is used to select a deflection link with the second highest Q-value in the Global Table. The experiment results show that, the three schemes above (RLDRS, RLAR, and IRLRCR) achieve lower burst loss probability than the SPF and Shortest Path Deflection Routing (SPDR) which selects the neighbor node in the second shortest path towards the destination as the deflection node. However, these three methods result in a larger mean burst end-to-end delay compared to SPF method \cite{NRL-LP}.

Haeri \emph{et al.} present a Predictive Q-learning algorithm for Deflection Routing (PQDR) in OBS \cite{PQDR} which is an extention of RLDRS \cite{RLDRS}. PQDR holds the idea that: the ability of packets processing of each link is different, and thus, the time is different for each link to recover from congestion status to normal status. The PQDR introduces the recovery rate to predict the link status for calculating Q-value in RLDRS. In PQDR, the Q-value update does not only consider the probability of successful transmission and delay, but also takes the packets processing ability of each link into consideration. The comparison shows that, PQDR outperforms the RLDRS in burst loss probability and deflection ratio, but has a larger average end-to-end delay. That is because PQDR has five lookup tables, which make the Q-value update process more complicated \cite{DRCN}.
\subsection{RWA and RSA in Optical Networks}
Besides routing resources assignment, allocation of other resources is also an important issue. Especially for the optical networks, there are multiple types of resources to be assigned, such as wavelength, spectrum, and modulation format, which makes the resource allocation more complicated. In WDM networks, several wavelengths are transmitted in a single link.  RWA problems refer to assign a route and an optical wavelength for each IP service flow \cite{RWA}, while RSA aims to setup a lightpath with multiple spectrum slots for a flow transmission. Spectrum assignment refers to allocate suitable spectrum slots to the requested lightpath. RSA in EON is the equivalent problem to the RWA in WDM networks. The difference of RSA and RWA is due to the capability of the EON architecture to offer flexible spectrum allocation to meet the requested data rates \cite{RSA}.

Both of RWA and RSA are NP-complete \cite{RSA-NP}. The problems are usually formulated as constrained optimization problems, such as Integer Linear Programming (ILP). Heuristics approaches, such as genetic algorithm and simulated annealing, are used to solve the problems \cite{RSA-ILP-GA} \cite{RSA-ILP-SA}. Recently, many researches leverage ML techniques to solve RWA and RSA problems. These works are introduced in this subsection.
\subsubsection{Supervised Learning-based RWA}
In \cite{MLRWA}, the RWA problem is modeled as a multi-class classification problem and solved with logistic regression and DNN. The optimal RWA configurations are computed in advance through ILP, and Net2Plan network optimization tool \cite{N2P} is used to obtain training samples. When training the logistic regression and DNN models, network states include topology, capacity, available wavelengths and the set of traffic demands, are input to the model, and the target output is the optimal RWA configuration to these states. The supervised learning model learns the relationships between network states and optimal optical RWA. With these learned relationships, computing the optimal solution with ILP is not necessary, which avoids high computational complexity. The computing of optimal RWA configuration by the above approach is time efficient and thus enable real-time network configuration.
\subsubsection{Reinforcement Learning-based RWA}
Kiran \emph{et al.} present a RL-based path and wavelength selection method in OBS networks to minimize the burst loss probability by learning an optimal path to an egress node over time \cite{RL-PSWS}. A RL agent is adopted at each ingress node. In path selection, since the learning function only installed in the edge nodes instead of intermediate nodes, there is no deflection routing in this method. The SPF method is first exploited as the initial path algorithm at the beginning, and the learning agent will select a route with the largest Q-value in each time epoch. In wavelength selection, since the presence of wavelength converters are not assumed at the core nodes, the wavelength selection agents are also deployed at ingress nodes. The wavelength selection agent will select the wavelength with the largest Q-value. The Q-values in routing selection and wavelength selection agents are updated depending on rewards corresponding to whether a burst is transmitted successfully. Experimental results show that the RL-based RWA method outperform the SPF, hybrid path switching scheme DWNV \cite{RL-PSWS2} and Self-Learning (SL) scheme  \cite{RL-PSWS6}. 

Pointurier \emph{et al.} consider the physical impairments in RWA \cite{RL-ROUTING}. In this method, each traffic should obey two constraints to avoid being blocked: wavelength continuity constraint and lightpath QoT constraint.  Wavelength continuity constraint is that, each traffic should occupy only one wavelength during its transmission and cannot be converted; lightpath QoT constraint is that, the QoT of a lightpath should reach a certain threshold to be used. In QoT evaluation, four physical impairments are considered: Inter-Symbol Interference, amplifier noise, inter-channel crosstalk, and node crosstalk, and BER is considered as the metric of the QoT. When a node choosing a lightpath with RL methods, the QoT of this lightpath is first examined and added into the candidate lightpath list only if it meets the QoT requirement. The method achieves a lower network blocking probability than the shortest path and uniform path selection schemes.
\subsubsection{Reinforcement Learning-based RSA}
In EON, the modulation format is also a reconfigurable parameter to meet the dynamic traffic requirements. Chen \emph{et al.} demonstrate routing, modulation format, and spectrum assignment in EON with DRL in \cite{DRMSA}. The action set of DRL agent are predefined resources assignment schemes. A deep Q-network is built to learn the best RMSA policies from actions set considering EON states (topology, connectivity, and spectrum utilization) and lightpath requests. The first few layers of Q-network are in the form of CNN, which responsible for extracting features from the network and traffic states. The latter full connection layers in Q-network compute the Q-value using the features learned by convolutional layers.
\section{Machine learning for Intelligent Optical Networks Monitoring and Survivability}
The awareness of the network states and performance, which provides essential information for network control and management, is important. The Optical Performance Monitoring (OPM) obtains the physical layer performance related to optical signal, optical links, and devices. It gives the basic performance parameters of degradation and impairments. After getting the underlying performance, QoT in lightpath level can be estimated to provide reference information to resource assignment and service restoration. Failures in optical networks may cause severe performance degradation in upper layer of communication networks, such as IP layer. Thus, detection and localization of failure is important for optical networks maintenance to improve survivability. In this section, use cases of machine learning in optical network monitoring and survivability are reviewed. Table 2 and Table 3 briefly review the important information of these use cases.
\subsection{Optical Performance Monitoring}
Optical networks are operated at ultra-high data rates, so that a short service interruption caused by impairment in fiber or devices may result in large-scale packet loss. Therefore, an accurate and real-time OPM is essential to ensure  network performance robust\cite{OPM-KILPER}.

The OPM techniques are used both in directly detected systems and digital coherent systems \cite{OPMCFT}. ML-based OPM are most used in directly detected systems where monitoring devices only use photodetectors to detect the intensity of the optical signal, or detect the electrical domain signal that transferred from the optical domain. 

Most prevalent parameters that OPM concerns include Optical Signal-to-Noise Ratios (OSNR), Chromatic Dispersion (CD), Polarization Mode Dispersion (PMD), polarization-dependent loss, optical power, and fiber nonlinearity \cite{OPMCFT}. The performance degradation of these parameters will cause signal impairments. 

In the following, a wide range of ML-based monitoring techniques for optical networks will be described.
\subsubsection{Neural Networks-based OPM techniques}
Jargon \emph{et al.} develop an ANN-based model for simultaneous estimation of OSNR, CD, and PMD using features that are extracted from eye diagrams \cite{ANN-OPM-EYE}. After the eye diagrams are captured, four features (Q-factor, closure, root-mean-square jitter, and crossing amplitude) are extracted from the diagrams. To estimate three parameters concurrently, the output of ANN is set to three neurons, with each neuron represents OSNR, CD, and Differential Group Delay (GDG, a parameter that reflects the level of PMD), respectively. The experiment results demonstrate the effectiveness of the model in the environment of 10 Gb/s NRZ-OOK and 40 Gb/s RZ-DPSK. 

The work in \cite{ANN-OPM-EYE} assumes the experiment systems are impairment-free in fibers itself, and do not consider the inherent impairments. Wu \emph{et al.} extend the work in \cite{ANN-OPM-EYE}, considering the impairments that experiment systems inevitably contain \cite{ANN-OPM}. In such a system with impairments, it is more practical for OPM to focus on the monitored parameters (e.g. optical power, OSNR, CD, and GDG) changes from a baseline, rather than the absolute values of these parameters. Furthermore, a method for estimating time misalignment in RZ-QPSK using ANN is provided in \cite{ANN-OPM}. 

The capture process of eye diagrams needs accurate clock recovery and corresponding complicated circuitries, which make monitoring more complex. Comparing to eye diagrams, Asynchronous Amplitude Histograms (AAHs) and asynchronous sampling diagrams do not need clock information and thus are more efficient to be used in OPM. 

Wu \emph{et al.} demonstrate a technique of ANN with features extracted from balanced-detected Delay-Tap Asynchronous Diagrams (DTADs) in 40 Gb/s RZ-BPSK system to estimate OSNR, CD, and DGD \cite{ANN-DTAD}. For PSK signals, a balance detection is used to produce more distinct features in waveforms to compensate the loss of phase information of the directly-detected signals. In \cite{OPM-ANN-BDAD}, a similar method is used in 100 Gb/s QPSK system and the correlation coefficients (metric for accuracy) are reported as 0.96 with directly-detection method, and 0.995 with balanced detection method.

The above ANN-based OPM techniques need manually extracted features as input which may require expert knowledge and extra labor. With the automated feature extraction ability of neural networks, especially DNN, the source data can be directly used as input of without specific feature engineering. 

Shen \emph{et al.} directly input the AAH diagram into the RBF neural network for OPM without extracting features from the graphs \cite{ANN-OPM-AAH}. The amplitude values and corresponding occurrences in histograms are used as the input of neural networks. OSNR, CD, PMD of optical signal are monitored in the modulation format of RZ-DQPSK and NRZ-16-QAM. The method does not require timing/clock recovery and thus are applicable to different modulation formats with different symbol rates. Tanimura \emph{et al.} demonstrate the use of DNNs for OSNR estimation without feature engineering \cite{OSNR-DNN-ASD}. The input data are 2048-dimension vectors, which are asynchronously sampled from the signal. The estimating results perform well for 16Gb/s DP-QPSK signal. The authors extend their works, and propose a CNN-based OSNR monitoring, using sampled data from signals \cite{CNN-OPM-OTN}. The sampled data are reshaped to a matrix and input as a graph to CNN model. Furthermore, authors investigate the kernels in CNN, with visualization tools and Fast Fourier Transform (FFT) tools, to explain what CNN has learned in the training phase. The results show that, CNN can learn to separate the signal from the noise by using bandpass-like filter. Wang \emph{et al.} introduce CNN for OSNR estimation \cite{CNNOSNREYE} with compressed eye diagrams. The eye diagrams are first transformed to grayscale and down sampled to a smaller graph. The preliminary processed eye diagrams are directly fed to CNN. The OSNR estimating with CNN perform better when compared with other ML algorithms (Decision Trees, KNN, BP-ANN, and SVM) \cite{CNNOSNREYE}.
\subsubsection{PCAs-based OPM}
To overcome the high computational complexity drawbacks when using images as inputs, Tan \emph{et al.} employ PCA to reduce the dimensionality of asynchronous delay-tap diagrams for efficiently estimating OSNR, CD, and DGD \cite{PCA-SOPM}. The ADTPs images are transformed to several Principal Components (PCs) with Karhunen-Loeve Transform (KLT). To estimate the impairment values, the PCs were compared with all the available feature vectors in the reference database, and the parameters values of reference data with minimum Euclidean distance are assigned to the unknown parameters. 
\subsection{Quality of Transmission Estimation}
In real optical networks operations, large system margin should be allocated to cover all uncertainties in the networks to guarantee reliable optical connectivity. These redundancies which may cause a large waste of network resources. An accurate monitoring of lightpath QoT can reduce the performance uncertainty, and thus reduce the redundant system margin to be allocated. 

To overcome the complexity and time-consuming computations of conventional analytic QoT estimation methods, ML methods have been used to learn relationship between network status and QoT from historical dataset. With ML-based accurate QoT estimation, capital expenditures (CAPEX) can be reduced.

Distinguished from section 4.1 that summarize the works monitoring the physical parameters of optical signals, this subsection emphasizes on estimating performance in the level of lightpath and channel, such as estimation of QoT of lightpath, and find relationships among channels and the links. 
\subsubsection{Lightpath QoT Estimation}
Barletta \emph{et al.} predict whether the Bit-Error-Rate (BER) of a lightpath meet the QoT requirement. This problem is modeled as a classification tasks, and random forest is employed as classifier \cite{QOT-ML}. The input features of the random forest include \emph{i)} the number of links of the lightpath; \emph{ii)} the total length of the lightpath; \emph{iii)} the length of longest link in the lightpath; \emph{iv)} the traffic volume a lightpath serves; \emph{v)}the modulation format. The predicted output is a binary variable which is True if a lightpath BER is lower than the system threshold. The authors evaluate several architectures of random forests and choose the classifier with 100 estimators, which provides the best trade-off between performance and computational time. The classification results can be used for reference when deploying a new lightpath in RSA.

Neural networks are exploited for estimating the blocking probability of bufferless OBS/OPS networks in \cite{ELM-BP}. Extreme Learning Machine (ELM) \cite{ELM-BP17} is a kind of neural network with a faster training speed because its input biases and input weights of the hidden nodes are randomly selected. Six input parameters of ELM including \emph{i)} mean traffic load in source-destination pair; \emph{ii)} difference between max and min traffic load in source-destination pair; \emph{iii)} number of channels per wavelength; \emph{iv)} number of wavelengths in networks; \emph{v)} average path length and \emph{vi)} concentration of route. The training set is the combination of variation of different values of six parameters and corresponding blocking probability.

A Case-based Reasoning (CBR) method is proposed to estimate the lightpath QoT and to classify the lightpaths into high- or low-quality categories in impairment-aware wavelength-routed optical networks \cite{CQ-CON}. The CBR \cite{CQ-CON14} method stores a knowledge base to enable the network exploiting previous experiences to solve the classification problems. In this work, CBR method stores a knowledge database with each sample consists of a set of attributes that describe a lightpath and corresponding Q-factor value. The attributes of a new lightpath will be compared with each sample in the knowledge base, and the weighted Euclidean distance is computed to evaluate similarity between two lightpaths. The corresponding Q-factor of the lightpath in the knowledge base with the largest similarity is assigned to the new lightpath as its Q-factor. The lightpath Q-factor is compared to a Q-factor threshold (Q-threshold) for lightpath performance classification. The authors also discuss a CBR with a learning and forgetting techniques to optimize the knowledge base to decrease its complexity.
\subsubsection{QoT Related to Channel Usage}
In the scenario of WDM networks, the lightpath QoT is not only related to attributes of lightpath itselt, such as length and link number, but also related to channel usage status. The channel ON/OFF states will affect the total link performance, such as crosstalk. Thus, different ON/OFF states correspond to different QoT performance, and ML is suitable to find relationship between them.

Samadi \emph{et al.} propose an ANN-based method to learn the relationship between the total OSNR value in a fiber with the ON/OFF states of WDM channels \cite{QOT-WDM}. The method does not require knowledge of fibers and Erbium Doped Fiber Amplifier (EDFA) specifications, only input with the ON/OFF states of the WDM channels, which is a 40-dimension 0-1vector in 40-channel WDM mesh network testbed. Two approaches are tested: \emph{i)} deploying one neural network for whole optical network, and \emph{ii)} one neural network for each node in network. Both approaches reach a relative low Root-Mean-Squared-Error (RMSE). 

The power excursion of EDFA in optical networks will degrade the QoT of lightpath in WDM systems. It is a channel-dependent effect that related to channel ON/OFF states. Identifying the relationship between power excursion with channel usage information can give recommends for channel establishment and energy pre-adjustment to improve lightpath performance. 

Huang \emph{et al.} map the channel usage states to power excursion values in multi-span EDFA links using Kernelized Bayesian Regression (KBR) \cite{ADD-DROP}. The authors define the Standard Deviation (STD) as the metric of power excursions level, and the main purpose of the estimation method is to lower the channels power STD. In the experiment environment with 24 DWDM channels, the input of the KBR model is a 24-bit array with each bit indicating an ON channel as 1, or an OFF channel as 0. The experiment shows that the STD prediction method achieves an accurate regression with low MSE.A further study investigates magnitude and correlation relationship between each channel state and EDFA power excursion, which is presented in \cite{DP-EDFA}. The magnitude that each channel contributes to total EDFA power excursion is fit by ridge regression.  And correlation means whether a rise in channel’s pre-EDFA power will increase or decrease the total EDFA power excursion, which is classified with logistic regression. The magnitude and correlation information are exploited for power adjustment to maintain power stability throughout the defragmentation process.
\subsection{Failure Management}
Failures in optical networks may cause network performance degradation and even huge data loss. There are mainly two kinds of failure management methods: reactive methods and proactive methods. In reactive methods, the network operators take actions when alarms occur. However, the data loss has already happened and the manual operation may not cope with massive alarms and invisible failures in aspects of flexibility and timeliness. Thus, the system should be designed in a proactive way, which can detect the potential failures and provide enough time for restoration actions \cite{MVN}. Machine learning is a powerful tool for proactive failure detection, which learns the relationship between current network status and future network failures. In addition to failure prediction, it is also nontrivial to identify and localize the root-causes of failures for efficient and precise service maintenance and restoration \cite{FMB}. In this subsection, the applications of ML techniques used in failures prediction, identification and localization will be reviewed.
\subsubsection{Failure Prediction}
Wang \emph{et al.} propose a failure prediction scheme with feature prediction and SVM \cite{FPML}. This work focuses on predicting the failures of boards in software defined metropolitan area network. The failure prediction is separated into two steps: \emph{i)} predict values of board performance indicators with Double Exponential Smoothing (DES), and \emph{ii)} detect failures with the predicted indicators. The historical values from \emph{t}-\emph{n} to time \emph{t}-1 of indicators are input to the DES algorithm, and the future values at \emph{t}+\emph{T} will be predicted. The kernel-based SVM model is trained with predicted values and classify whether the failure will occur at \emph{t}+\emph{T}. Besides, the correlationbetween indicators and board failure is calculated and several SVM model with different kernels is compared. 
\subsubsection{Failure Identification}
Rafique \emph{et al.} propose a transport SDN-integrated (TSDN) cognitive assurance architecture with ANN-based proactive fault detection \cite{CAA}. To simplify the computational complexity of neural networks, the received optical power is pre-tested with generalized Extreme Studentized Deviate (ESD) test, and only indicators of the tests are input into the neural networks. The pre-test procedure takes the load off the neural networks. The output of neural network suggests true normal or true abnormal behavior of a potential identified failure. The experiment is carried on a diverse pattern extracted from ADVA sample network \cite{CAA25}, and the proposed assurance architecture is compared with the threshold-based failure detection method. Experimental results show that the ANN-based proactive detection outperforms the condition-based reactive detection both in detection accuracy and response time (defined as time from detection to typical failure state).

Two kinds of filter-related soft failures, i.e. filter shift and filter tightening, are identified with decision trees and SVMs \cite{LFOS}. The spectrum of a signal may become asymmetric after passing the filter with \emph{filter shift} failure, and spectrum edges get noticeably rounded with \emph{filter tightening} failure. All these changes can be monitored in Optical Spectrum Analyzers (OSAs); thus, these real-time OSA monitoring results can be used for identification of the root causes of the failures. Three models \emph{i)} multi-classifier approach, \emph{ii)} single-classifier approach, and \emph{iii)} residual-based approach are proposed to classify the failures. In multi-classifier approach, classifier is used at each intermediate node, and the number of filters is taken into consideration. In single-classifier training, the filter cascade effect should be masked in feature engineering stage. The signal should be compensated by adding/subtracting the differences between properly configured signal before feature extraction. In residual-based approach, the signal is compared with an ideal signal that pass the same number of filters without failures. The differences between them are input into the ML models for classification. 

Ruiz \emph{et al.} identify failures using Bayesian Networks (BN) \cite{STFI}. Two kinds of failures, tight filtering and inter-channel interference, are considered. With the information of pre-FEC BER and received power $P_{\rm{RX}}$ (minimum, maximum, and average power), the BN gives the probability of whether there is a failure in the link and what kind of failure it is. Then, the system reconfigures the lightpath to solve the BER degradation. 
\subsubsection{Failure Localization}
With increasing number of network elements, there will be massive failure alarms in optical networks. Due to the complexity of network topology and the connectivity of network components, several alarms may be trigger by only one failure. Zhao \emph{et al.} propose a massive alarm analyzing method with Deep Neural Evolution Networks (DNEN) to accurately localizing failure in WDM networks \cite{AFL}. The proposed method extracts the deep hidden failure features and localizes the real failures. DNEN generates a series of neural networks, and create new networks by crossover and mutation until the model meet the accuracy requirements. DNEN generates a series of neural networks, and use crossover and mutation among these neural networks instead of gradient descent for training to jump out of local optimum. The training data come from real optical networks, and experimental results show that the DNEN-based method can achieve the highest failure localization accuracy compared to SVM and DNN based methods. Besides, it is time efficient and suitable for practical using.

Vela \emph{et al.} propose a data visualization method for failure localization with the aid of K-means \cite{ADVFL}. To support human operation with monitored path BER data, the BER performance of paths and their change trends are visualized.  Paths are clustered according to the max BER and BER trend with K-means approach, and different centroids are plotted in a 2D place. Different colors are assigned according to different BER performance for better representation to human operators.

\section{Challenges and possible Solutions}
Although applying ML for intelligent optical networks has achieved better efficiency and accuracy than many conventional methods, there still exists several challenges to be solved. In this section, challenges of applying ML for intelligent optical networks and possible solution will be discussed. 
\subsection{Open Dataset Access}
Open datasets are crucial to applying ML because the performance of different methods can be compared based on the same dataset. In this way, an easily accessible open dataset will reduce the repetitive works and accelerate academic research progress. However, in the field of optical networks, there only a few open datasets available. Most works that previously surveyed in this paper use synthetic data for ML model training, but may lack credibility, and is difficult to be compared with other works.

There are several hinders in collecting and opening dataset of the real optical networks. Firstly, in real network operations, large resource margins will be reserved to maintain a good performance. Therefore, there are few negative samples in real networks, and the datasets collected from the real environment may suffer data imbalance problem that the positive samples are much more than negative samples. Secondly, the real telecommunication data may face privacy issues, which make it difficult for Internet Service Providers (ISPs) to make dataset public.

\textbf{Possible solutions:}
\emph{i)} Standardize the process of data collecting, labeling, cleaning and anonymizing to reduce the costs of data processing, and keep the data with privacy. \emph{ii)} Train the ML models with encryption mechanism, such as federated learning algorithm proposed in 2016 \cite{FL}, with which models can be trained without exact access to the data, and thus privacy can be guaranteed. Such method has been used in communication systems \cite{AFLR}.

\subsection{Model Interpretability and Traceability}
In operation and maintenance tasks of optical networks, monitoring results should be interpretable to operators, and configuration actions should have clear reasons to be taken. However, most ML algorithms work in a black-box way. Although their training processes are open and transparent, the trained models are uninterpretable. Without the interpretability, it may be difficult for network operators to troubleshoot the problems when network performance is not as good as expected.

In addition to interpretability, ML also lacks traceability, which focuses on tracing from output results back to input features. For example, a neural network model is used for optical device failure prediction, with input parameters include ambient temperature and usage time of the device. When the device is predicted to break down, it is not sure if it is because the ambient temperature is too high or the device has reached the end of its service life. Tracing the cause of failures is as important as finding them, because different failure reasons correspond to different operation actions. 

\textbf{Possible solutions:}
\emph{i)} More interpretable models, such as logistic regression and tree-based ML model, should be employed with priority \cite{IML}. It is undeniable that, these models may be too simple to handle the complex problems in optical networks; however, it is still worthwhile to try these models first to find tradeoff between interpretability and performance. \emph{ii)} Seek a balance between rule-based and data-driven methods. In optical networks, there have been existing explicitly known relationship and expert knowledge, so it is possible to exploit these relationships as built-in rules to construct learning models in a ``top-down'' approach, in which the whole structure of problem-solving is based on human knowledge, and ML only acts as submodule, such as value-fitting. With these built-in rules, the interpretability of a ML model will be promoted.
\subsection{Model Generalization Ability}
Generalization ability is an important measure to evaluate a ML model, and is especially important when applying ML for intelligent optical networks environment. Because in the network problems, the trained ML model is tightly coupled with the network environment. The model can only target one network structure or scenario. If the network environment changes, the model retraining is necessary, which will be of great computational cost.

\textbf{Possible solutions:} 
\emph{i)} Decouple the input features and network state in the feature engineering process, for example, training the model with features focus on the local network state instead of using features only focus the total network. \emph{ii)} Adopt algorithms that are capable of online learning, which can fine tune the model itself, instead of retraining when the environment changes.
\subsection{Algorithm Computational Complexity}
In real optical network scenario, there may be strict requirement for the timeliness of tasks. Thus, it is important for ML algorithms to reduce their computational complexity. However, only a small set of works analyze the ML feasibilities based on time complexity. 

Besides, most of the previous experiments demonstrate the effectiveness of the proposed methods without time restriction. These optimal results were obtained assuming there is enough time for computation. However, real environment is usually time-sensitive, and a better comparison principle is to compare the suboptimal solutions of different algorithms under a given time threshold.

\textbf{Possible solutions:}
\emph{i)} In monitoring tasks, although many ML algorithms, such as deep neural networks, have the capability of automated feature extraction, the original data still should be preprocessed and expert knowledge should be introduced into feature engineering to simplify the model training, e.g. pre-test the data, and only input the indicators of tests into neural networks for training \cite{CAA}. \emph{ii)} In decision-making tasks, suboptimal solutions should be allowed to balance the network performance and time consuming of action computation.

\subsection{System Security and Reliability}
The adoption of ML will increase the flexibility and automaticity of optical network and reduce the necessity of manual operations. However, ML models often work in a best-effort way, and do not provide performance guarantee, which may cause security and reliability issues. Security issues refer to inherent vulnerability in ML models, and reliability issues refer to performance degradation or errors of trained ML models. The lack of security and reliability guarantee may hinder the practical use of ML in real networks.

\textbf{Possible solutions:}
\emph{i)} Establish periodical model effectiveness evaluation mechanism and model performance degradation alarm mechanism. \emph{ii)} In the system design stage, ML-aided mode is suggested to adopted instead of ML-dominate mode, and interfaces for manual intervention need to be reserved.

\subsection{Reality Gap Between Simulation and Real Network }
Most ML methods are demonstrated in a simulation environment or in small scale network, very few have been demonstrated in real optical network. It is difficult to test ML methods in the real network due to security and privacy concerns. However, how much characteristics of a real network can be emulated, and whether the model performance in emulation network is as same as in real network, are difficult to evaluate. These reality gaps may hinder the well-trained ML model under a simulation platform being used directly in real networks. 

\textbf{Possible solutions:}
\emph{i)} Take more characteristics of real networks (both physical parameters and network effects) into consideration when building simulation platforms, to narrow the reality gap. \emph{ii)} Although putting methods into practice is costly, more field tests are still worthy to be taken to evaluate the proposed methods in real networks.
\section{Conclusion}
As an inter-disciplinary blend, applying ML in optical networks is a promising approach to improve the performance of optical networks. In this paper, existing researches that apply ML for intelligent optical networks are reviewed. We begin our discussion with background and challenges of current communication networks and optical networks. Thereafter, three paradigms of ML (regression, classification, and decision-making) are discussed in detail. Motivations of using ML to build intelligent optical networks are analyzed from aspects of both inherent characteristics of ML algorithms and external enabling techniques. Then, ML algorithms which have been using in optical networks are reviewed. In the main part of this survey, various ML use cases in optical networks are reviewed from two categories, include optical network control and resource management, and optical network monitoring and survivability. Finally, we identify challenges that may be faced by future applications of ML for intelligent optical networks, and propose possible solutions to each challenge. 

In summary, it is nontrivial to introduce intelligence into optical networks, and exploration of applying ML in intelligent optical network is on its rise. This paper attempts to investigate how ML techniques have been and should be used in optical networks. We hope that our discussion and exploration may provide convenient to researchers to apply ML in future intelligent optical networks.

\bibliography{mybibfile}
\newcommand{\tabincell}[2]{\begin{tabular}{@{}#1@{}}#2\end{tabular}}  
\begin{center}
\begin{table}
\caption{Applications of ML for intelligent optical networks control and resource management}
\begin{tabular}{ccccc}
\hline
Work  &  \tabincell{c}{ML \\ Algorithms}  &  Task  &  Input data  &  Network Scenario  \\
\hline
\multicolumn{5}{c}{\textbf{Traffic and Resource Requirement Prediction}}  \\
\hline
\cite{DLID}  &  DNN  &  \tabincell{c}{Resource requirement \\ prediction}  &  Historical traffic  &  \tabincell{c}{Inter-datacenter \\ optical networks}  \\
\hline
\cite{LDLA}  &  DNN  & \tabincell{c}{Traffic arrival time and \\ resource prediction}  &  Historical traffic  &  \tabincell{c}{Intra-DCN, \\ real datacenter}  \\
\hline
\cite{HIDCN}  &  NARNN  &  Traffic prediction  &  Historical traffic  &  Intra-DCN  \\
\hline
\cite{MLBP}  &  LSTM  &  Traffic prediction  &  Historical traffic  &  \tabincell{c}{Intra-ODCN, \\ NSFNET topology}  \\
\hline
\multicolumn{5}{c}{\textbf{Routing in Optical Networks}}  \\
\hline
\tabincell{c}{GRPM \\ \cite{GPRM-OBS}}  &  \tabincell{c}{Bayesian \\ Network}  &  \tabincell{c}{Routing}  &  \tabincell{c}{Offset time, burst loss ratio, \\ number of hops, destination}  &   \tabincell{c}{NSFNET, Network\\Simulator 2 (ns-2)} \\
\hline
\cite{MLAR}  &  \tabincell{c}{Logistic \\ Regression}  &  Routing  &  Traffic matrices  &  \tabincell{c}{12 OpenvSwitches and \\ 4 end hosts, ONOS \\ SDN controller}  \\
\hline
\cite{WITHOUT}  &  LSTM  &  Routing  &  \tabincell{c}{Traffic requests and \\ inter-domain link capacities}  & \tabincell{c}{Multi-domain network \\with  nine domains} \\
\hline
\tabincell{c}{RLAR \\ \cite{NRL-LP}}  &  RL  &  \tabincell{c}{Routing in OBS}  &  ------  &  \tabincell{c}{NSFNET, regular \\  4*4 torus topology}  \\
\hline
\cite{MARL-OBS}  &  \tabincell{c}{Multi-agent \\ RL}  &  \tabincell{c}{Routing in OBS}  &  ------  &  \tabincell{c}{NSFNET, \\ random topology}  \\
\hline
\tabincell{c}{RLDRS \\ \cite{RLDRS}}  &  RL  &  \tabincell{c}{Deflection routing}  &  ------  &  \tabincell{c}{NSFNET, regular \\  4*4 torus topology}  \\
\hline
\tabincell{c}{IRLRCR \\ \cite{NRL-LP}}  &  RL  &  \tabincell{c}{Deflection routing}  &  	------  &  \tabincell{c}{NSFNET, regular \\  4*4 torus topology}  \\
\hline
\tabincell{c}{PQDR \\ \cite{PQDR}}  &  RL  &  \tabincell{c}{Deflection routing}  &  ------  &  \tabincell{c}{NSFNET, \\ Waxman topologies}  \\
\hline
\multicolumn{5}{c}{\textbf{RWA and RSA}}  \\
\hline
\cite{MLRWA}  & \tabincell{c}{ Logistic Regression,\\ DNN}  &  RWA  &  \tabincell{c}{Network features \\ and traffic matrices}  &  \tabincell{c}{5-node Spanish network \\topology, Abilene topology}  \\
\hline
\cite{RL-PSWS}  &  RL  &  RWA  &  ------  &  \tabincell{c}{OBS, 14- and 21-node \\NSFNET, random topology, \\ns-2 simulator}  \\
\hline
\cite{RL-ROUTING}  &  RL  &  \tabincell{c}{RWA}  &  ------  &  NSFNET \\
\hline
\cite{DRMSA}  &  DRL  &  RMSA  &  Link and spectrum usage table  &  Six-node EON topology  \\
\hline

\end{tabular}
\end{table}
\end{center}

\begin{landscape}

\begin{table}
\caption{Applications of ML for intelligent optical networks monitoring and survivability: Optical Performance Monitoring}
\begin{tabular}{cccccc}
\hline
Work  &  \tabincell{c}{ML \\ Algorithms}  &  \multicolumn{2}{c}{Input data}  &  Task  &  \tabincell{c}{Experiment \\Environment}  \\
\hline
\multicolumn{6}{c}{\textbf{Optical Performance Monitoring}}  \\
\hline
\quad	&                \quad                             &  Input Source Data  &  Input Features       &     \quad                 &\quad \\
\hline
\cite{ANN-OPM-EYE}  &  ANN  &  \tabincell{c}{Eye \\ diagrams}  &  \tabincell{c}{Q-factor, closure, \\ jitter, crossing-amplitude}  &  \tabincell{c}{OSNR, CD, DGD estimation}  &  \tabincell{c}{10 Gb/s NRZ-OOK, \\40 Gb/s RZ-DPSK}  \\
\hline
\multirow{2}{*}{\cite{ANN-OPM}}   &  \multirow{2}{*}{ANN}  &   \multirow{2}{*}{\tabincell{c}{Eye \\ diagrams}}  &  \tabincell{c}{Q-factor, closure, jitter, \\ crossing-amplitude, the \\mean and SD of '1's and '0's}  &  \tabincell{c}{Changes in optical power, \\OSNR, CD, and GDG estimation}  &  \tabincell{c}{40 Gb/s NRZ-OOK \\and RZ-DPSK}  \\  \cline{4-6}
			&  &  &  \tabincell{c}{RF clock tone power, \\Low-frequency RF power}  &  \tabincell{c}{Time misalignment estimation}  &  DQPSK \\
\hline
\cite{ANN-DTAD}  &  ANN  &  \tabincell{c}{Balanced detected \\DTADs}  &  \tabincell{c}{Magnitudes features}  &  \tabincell{c}{OSNR, CD, PMD estimation}  &  40 Gb/s RZ-BPSK   \\
\hline
\cite{OPM-ANN-BDAD}  &  ANN  &  \tabincell{c}{Balanced detected \\Asynchronous Diagrams}  &  \tabincell{c}{Magnitudes features}  &  \tabincell{c}{OSNR, CD, PMD estimation}  &  100 Gb/s QPSK   \\
\hline
\cite{ANN-OPM-AAH}  &  \tabincell{c}{MIMO \\RBF NN}  &  AAHs  &  \tabincell{c}{Amplitude levels and \\corresponding occurrences}  &  \tabincell{c}{OSNR, CD, PMD estimation}  &  \tabincell{c}{40 Gb/s RZ-DQPSK, 40 Gb/s 16-QAM}   \\
\hline
\cite{OSNR-DNN-ASD}  &  DNN  &  Sampled signal  &  HIs, HQs, VIs, VQs  &  OSNR estimation  &  16 Gb/s DP-QPSK   \\
\hline
\cite{CNN-OPM-OTN}  &  CNN  &  Sampled signal  &  \tabincell{c}{HIs, HQs, VIs, VQs \\(in the format of matrices)}  &  OSNR estimation  &  DP-QPSK, 16QAM, 64QAM   \\
\hline
\cite{CNNOSNREYE}  &  CNN  &  Eye diagrams  &  Pixels of eye diagrams  &  OSNR estimation  &  \tabincell{c}{4PAM, RZ-DPSK, NRZ-OOK, RZ-OOK}   \\
\hline
\cite{PCA-SOPM}  &  PCA  &  ADTPs  &  \tabincell{c}{Principal components that \\extracted from ADTPs}  &  \tabincell{c}{OSNR, CD, DGD estimation}  &  \tabincell{c}{RZ-OOK, PM-RZ-QPSK, \\PM-NRZ-16QAM}   \\
\hline
\end{tabular}
\end{table}

\begin{table}
\caption{Applications of ML for intelligent optical networks monitoring and survivability: QoT Estimation and Failure Management}
\begin{tabular}{ccccc}
\hline
Work  &  \tabincell{c}{ML \\ Algorithms}  &  Input data  &  Task  &  \tabincell{c}{Experiment \\Environment}  \\
\hline
\multicolumn{5}{c}{\textbf{Quality of Transmission Estimation}}  \\
\hline
\cite{QOT-ML}  &  \tabincell{c}{Random Forest}  &   Lightpath features  &  \tabincell{c}{BER Estimation}  &  \tabincell{c}{NSFNET/Japan network topology, \\six modulation formats}   \\
\hline
\cite{ELM-BP}  &  \tabincell{c}{Extreme Learning Machine}  &   \tabincell{c}{Network features}  &  \tabincell{c}{Blocking Probability}  &  \tabincell{c}{OBS/OPS, 13-node NSFNET}   \\
\hline
\cite{CQ-CON}  &  \tabincell{c}{Case-based \\Reasoning}  &   Lightpath features  &  Q-factor  &  \tabincell{c}{Deutsche Telekom network, \\GEANT2 network, \\32/64wavelength per link, 10 Gb/s OOK}   \\
\hline
\cite{QOT-WDM}  &  Neural Network  &   ON/OFF status of channels  &  \tabincell{c}{Channel related \\QoT estimation}  &  \tabincell{c}{Mesh network testbed \\with 40 WDM channels}   \\
\hline
\cite{ADD-DROP}  &  \tabincell{c}{Kernelized Bayesian \\Regression}  &   ON/OFF status of channels  &  \tabincell{c}{Channel related \\QoT estimation}  &  \tabincell{c}{2 span/3 span networks,\\ 24WDM channels}   \\
\hline
\cite{DP-EDFA}  &  \tabincell{c}{Ridge Regression, \\Logistic Regression}  &  ON/OFF status of channels  &  \tabincell{c}{Relationship between power \\discrepancy and channel status}  &  \tabincell{c}{3 span networks, \\24WDM channels}   \\
\hline
\multicolumn{5}{c}{\textbf{Failure Management}}  \\
\hline
\cite{FPML}  &  SVM	  &  \tabincell{c}{Predicted indicators \\value with DES}  &  	Failure prediction  &  	\tabincell{c}{Software-Defined Metropolitan \\Area Network (SDMAN)}  \\
\hline 
\cite{CAA}	  &  Neural Network  &  	ESD test values  &  	Failure identification  &  	\tabincell{c}{Transport Software \\Defined Network (TSDN)}  \\
\hline
\cite{LFOS}	  &  \tabincell{c}{Decision Tree, SVM}  &   \tabincell{c}{Features extracted \\from optical spectrum}  &  	Failure identification  &  	VPI software simulation network  \\
\hline
\cite{STFI}  &  	Bayesian Network	  &  \tabincell{c}{Received power \\and Pre-FEC BER}  &  	\tabincell{c}{Failures localization \\and identification}	  &  \tabincell{c}{Nyquist Wavelength Divisioin \\Multiplexing (NWDM), 120 Gb/s QPSK}  \\
\hline
\cite{AFL}	  &  \tabincell{c}{Deep Neural Evolution\\Network (DNEN)}  &  	Alarm sets	  &  Failure localization	  &  Software defined optical network  \\
\hline
\cite{ADVFL}	  &  \tabincell{c}{K-means}  &  	Max BER and BER trend	  &  Failure localization	  &  \tabincell{c}{30-node, 50-link Spanish \\Telefonica optical network}  \\
\hline

\end{tabular}
\end{table}

\begin{table}
\caption{A summary of ML usage for intelligent optical networks}
\begin{tabular}{ccccccccc}
\hline
\multicolumn{2}{c}{\multirow{2}{*}{}} & \multicolumn{3}{c}{\tabincell{c}{Network Control and Resource Management\\}} & & \multicolumn{3}{c}{\tabincell{c}{Network Monitoring and Survivability \\}} \\  \cline{3-5} \cline{7-9}
&&Traffic Prediction & Routing & RWA and RSA & &OPM & QoT & \tabincell{c}{Failure Management}\\
\hline
\multirow{10}{*}{\tabincell{c}{Supervised\\ Learning}} & SVM & & & & &\checkmark & & \checkmark \\ \cline{2-9}
                                      & Logistic Regression &&\checkmark&\checkmark&&&\checkmark& \\ \cline{2-9}
                                      & NN &\checkmark&&&&\checkmark&\checkmark&\checkmark \\ \cline{2-9}
                                      & DNN &\checkmark&&\checkmark&&\checkmark&&\checkmark \\ \cline{2-9}
                                      & CNN &&&&&\checkmark&& \\ \cline{2-9}
                                      & LSTM &\checkmark&\checkmark&&&&& \\ \cline{2-9}
                                      & Case-based Reasoning &&&&&&\checkmark& \\ \cline{2-9}
                                      & Random Forest &&&&&&\checkmark& \\ \cline{2-9}
                                      & Decision Tree &&&&&\checkmark&&\checkmark \\ \cline{2-9}
                                      & Bayesian Network &&\checkmark&&&&&\checkmark \\ \cline{2-9}
\hline
\multirow{2}{*}{\tabincell{c}{Unsupervised\\ Learning}} & PCA & & & && \checkmark & &  \\ \cline{2-9}
                                       & K-means & & & & & & &\checkmark  \\ \cline{2-9}
\hline
\multirow{2}{*}{\tabincell{c}{Reinforcement \\Learning}} & RL & & \checkmark & \checkmark & & & &  \\ \cline{2-9}
                                       & DRL & & &\checkmark  & & & &  \\ \cline{2-9}

\hline

\end{tabular}
\end{table}

\end{landscape}
\end{document}